\documentclass[reprint,
superscriptaddress,
 amsmath,amssymb,
 aps,
]{revtex4-2}

\usepackage{graphicx}% Include figure files
\usepackage{dcolumn}% Align table columns on decimal point
\usepackage{bm}% bold math
\usepackage{color}
\usepackage{color}

\begin{document}

\preprint{APS/123-QED}

\title{Condensed Spin Excitation of Quantized Dirac Fermions in Quasi-Two-Dimensional Semimetal BaMnBi$_2$}% Force line breaks with \\

\author{Masashi Kumazaki}
\affiliation{Department of Physics, Nagoya University, Chikusa, Nagoya 464-8602, Japan.}%Lines break automatically or can be forced with \\
\author{Azimjon Temurjonov}
\affiliation{Department of Physics, Nagoya University, Chikusa, Nagoya 464-8602, Japan.}%Lines break automatically or can be forced with \\
\author{Takaaki Jinno}
\affiliation{Technical center, Nagoya University, Chikusa, Nagoya 464-8602, Japan.}
\author{Yukihiro Watanabe}
\affiliation{Department of Physics, Nagoya University, Chikusa, Nagoya 464-8602, Japan.}%Lines break ¹automatically or can be forced with \\
\author{Taku Matsuhita}
\affiliation{Department of Physics, Nagoya University, Chikusa, Nagoya 464-8602, Japan.}%
\author{Yoshiaki Kobayashi}
\affiliation{Department of Physics, Nagoya University, Chikusa, Nagoya 464-8602, Japan.}%
\author{Yasuhiro Shimizu}%
 %\email{yasuhiro@iar.nagoya-u.ac.jp}
\affiliation{Department of Physics, Nagoya University, Chikusa, Nagoya 464-8602, Japan.}
\affiliation{Department of Physics, Shizuoka University, Suruga, Shizuoka 422-8529, Japan.}%Lines break automatically or can be forced with \\ 
 
\date{\today}% It is always \today, today,

\begin{abstract}
Dirac semimetals provide a new platform for the quantum Hall effect at low magnetic fields. In the presence of strong spin-orbit coupling, a spin-split Landau level is expected to enhance the bulk quasiparticle excitation. Here we report NMR spectroscopy that site-selectively probes dynamic spin susceptibility on the magnetic semimetal BaMnBi$_2$. We find that spontaneous staggered fields from antiferromagnetic Mn moments are completely canceled at the Bi layer hosting Dirac fermions. The nuclear spin-lattice relaxation rate $1/T_1$ follows the cubic temperature dependence down to low temperatures under the in-plane field, manifesting the chemical potential close to the Dirac point. $1/T_1$ becomes constant below 20 K under the out-of-plane field, where the well-separated Laudau level appears. The strong anisotropy of $1/T_1$ exceeding 100 suggests spin-split Landau levels in the quantum Hall regime. 
\end{abstract}

\maketitle

The quantum Hall effect (QHE) with a chiral edge state protected by the bulk energy gap has been studied in two-dimensional (2D) electron systems, such as semiconductor interfaces requiring extremely high mobility under ultralow temperatures and high magnetic fields \cite{vKlitzing1980, Tsui1982, vonKlitzing2020}. The main probe of QHE has been limited in transport measurements and surface-sensitive spectroscopy \cite{Barrett1996, Kronmuller1999, Hashimoto2002NMR, li2007graphite, Melcer2024}, while fractional quasiparticles of composite fermions under electron correlation remain uncovered in the bulk state \cite{Moore1991, Wen1995}. Recent developments of relativistic Dirac semimetals with low carrier density and small effective mass realize the observation of the half-integer QHE at high temperatures and low fields \cite{Novoselov2007, Bolotin2009, Castro2009, He2014, Jeon2014,li2015graphene, Uchida2017, Kealhofer2020topological, Liu2022}. The semimetals with symmetry protected Dirac points allow one to investigate the bulk spin excitation of QHE \cite{Young2015, Tajima2013, masuda2016QHE, Tang2019ZrTe5}. 2D Dirac semimetals involve massless Dirac fermions that follow the Hamiltonian $H = v_{\rm F}(\sigma_x p_x + \sigma_y p_y )$ using the Fermi velocity $v_{\rm F}$, the Pauli matrix $\sigma_i$ of pseudospin, and the momentum $p_i$. The cyclotron energy is expressed as $\pm v_{\rm F}\sqrt{e\hbar|n|B}$ with the non-negative integer Landau index $n$ under the magnetic field $B$ \cite{Castro2009}. The four-fold degeneracy of the relativistic Landau level is lifted via spin-orbit coupling and Zeeman interaction. The quasiparticle excitation that accompanies spin-flip process can be strongly enhanced when the chemical potential crosses the spin-polarized Landau levels \cite{Goerbig2011}. In the absence of spatial inversion, the momentum-dependent spin-split valley leads to a spin-momentum locking useful in spintronics \cite{Lv2021}.

\begin{figure}
\includegraphics[width=8cm]{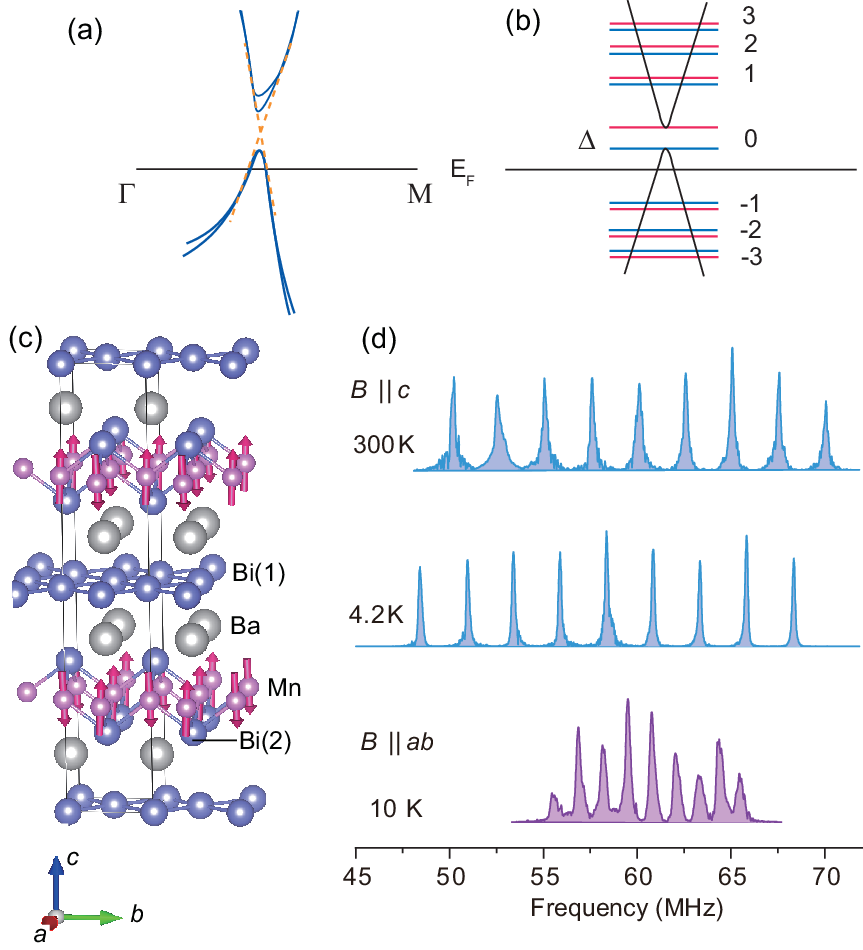}
 \caption{\label{FIG1}(a) Dirac cone along the $\Gamma$-M line in BaMnBi$_2$ \cite{ryu2018Zn}. (b) Spin-split Landau levels under spin-orbit and Zeeman coupling for one of the four valleys. The red and blue lines denote energy levels with opposite spins. (c) Crystal structures of BaMnBi$_2$ (tetragonal $I4/mmm$), including two Bi sites: Bi(1) forms a square-net layer on the mirror plane \cite{li2016Bi, ryu2018Zn}, and Bi(2) bridges localized Mn ions. The arrow on Mn denotes the magnetic moment below $T_{\rm N}$ \cite{lee2013Sr}. (d) $^{209}$Bi ($I=9/2$) NMR spectra under the out-of-plane (${\bf B} \parallel {\bf c}$) and in-plane (${\bf B} \parallel ab$ plane) magnetic field.}
\end{figure}

The semimetals $A$Mn$X_2$ ($A$: alkali- or rare-earth element; $X$: Sb, Bi) provide a playground for half-integer QHE in 2D Dirac fermions \cite{masuda2016QHE, Liu2016Sb, Huang2017Sb, ryu2018Zn, borisenko2019Yb, Klemenz2019review}. Among them, BaMn$X_2$ ($X$ = Bi, Sb) satisfies a feature of Dirac semimetals with a small effective mass $m^*\simeq 0.1m_0$ and a phase shift $\pi$ of the Shubnikov-de Haas (SdH) oscillation \cite{li2016Bi, Huang2017Sb, sakai2020Sb, Liu2021Sb, Kondo2021Bi}. Here, we focus on BaMnBi$_2$ expected to involve a single Dirac cone and strong spin-orbit coupling for investigating the spin-orbital entangled excitation of Dirac fermions [Fig. \ref{FIG1}(a,b)]. The tetragonal crystal structure ($I4/mmm$) consists of the square-net Bi$(1)$ layer that hosts Dirac fermions and the Mott insulating MnBi(2) layer with a localized Mn moment ($S = 5/2$), as shown in Fig. \ref{FIG1}(c) \cite{Ma2019, ryu2018Zn, lee2013Sr}. The local moment exhibits long-range antiferromagnetic order near room temperature ($T_{\rm N}$ = 290 K)\cite{li2016Bi}, resulting in the confinement of orbital motions into the square-net layer under the magnetic field perpendicular to the layer. There are four valleys of anisotropic Dirac cones along the $\Gamma$-M direction of the Brillouin zone for the $I4/mmm$ lattice [Fig. \ref{FIG1}(a)]. The SdH oscillation and Fermi surface depend sensitively on chemical potential and lattice distortion \cite{li2016Bi, ryu2018Zn, Kondo2021Bi}. The predominant hole carrier of the Dirac cone was observed in photoemission spectroscopy without lattice distortion \cite{ryu2018Zn}, whereas the negative Hall resistance suggests the Fermi level across the parabolic electron band under orthorhombic distortion \cite{Kondo2021Bi}.  

Nuclear magnetic resonance (NMR) is a powerful tool for low-lying excitations in Dirac and Weyl semimetals \cite{hirata2017I3, nisson2013nuclear, yasuoka2017Weyl, Tian2019, wang2020As, Tian2021ZrTe5, Papawassiliou2020, Watanabe, Hirata2021, Yokoo2022}, where the nuclear spin-lattice relaxation rate $1/T_1$ obeys a power law in temperature ($\propto T^3$) \cite{hirosawa2017T1, dora2009graphene, Katayama}. The Knight shift and $1/T_1$ due to the hyperfine interaction between Dirac fermions and nuclear spins are distinct from conventional metals, depending on the dimensionality of the system \cite{dora2009graphene, okvatovity2019TaP, Maebashi2019}. $1/T_1$ is sensitive to the location of chemical potential $\mu$ with respect to the discrete Landau levels. 

Here we investigate the local static and dynamic susceptibility through the $^{209}$Bi Knight shift and $1/T_1$, respectively, on the square-net Bi layer with on-site Dirac fermions in BaMnBi$_2$. The NMR spectrum reflects the local symmetry across long-range magnetic ordering, while the magnetic susceptibility is governed by Mn moments. $1/T_1$ measures spin fluctuations of Dirac fermions at low temperatures, where the anisotropy is sensitive to the degeneracy of Landau levels. We compare the results with a numerical calculation for quantized Dirac fermions and discuss the effect of spin-orbit coupling. 

Single crystals of BaMnBi$_2$ were prepared by a flux method in a sealed quartz tube \cite{chen2017Bi, Klemenz2019review, Kondo2021Bi}. The typical size of the crystal was $3\times 3 \times 0.5$. Magnetization was measured for a single crystal with a superconducting quantum interference device at 5.0 T. $^{209}$Bi NMR measurements were conducted on the single crystal ($^{209}$Bi nuclear spin $I=9/2$, nuclear gyromagnetic ratio $\gamma_{\rm n}$ = 6.841 MHz/T). Spin echo signals were taken by a $300$-kHz frequency step using the pulse sequence $t_{\pi/2}-\tau-t_{\pi/2}-\tau$ ($t_{\pi/2}$ = 1 $\mu$s, $\tau = 4-10$ $\mu$s) in a steady magnetic field of $B = \mu_0H$ = 8.51 T. The spin-echo decay time $T_2$ was located around 10 $\mu$s (250 K) and 50 $\mu$s (100 K). The nuclear spin-lattice relaxation time $T_1$ was obtained from a saturation recovery method for the central resonance line \cite{SM, Wada}. 

The $^{209}$Bi NMR spectrum of BaMnBi$_2$ consists of a set of nuclear quadrupole splits for $I=9/2$ [Fig. \ref{FIG1}(d)]. The observed spectrum comes from the Bi(1) site of the square net layer, while the spectrum from Bi$(2)$ bonded to Mn is wiped out due to the fast $T_2$ ($<1 \mu$s). Even below $T_{\rm N}$, no spectral split was observed for the magnetic field along the $c$ axis and the $ab$ plane. This means that the staggered local fields from the antiferromagnetic Mn moments are completely canceled out at the Bi(1) layer and only the uniform field contributes to the spectral shift. The result is compatible with the antiparallel spin structure along the $c$ axis with the mirror plane on the Bi(1) layer \cite{li2016Bi, ryu2018Zn}, as shown in Fig. 1(c). The spin structure is identical to SrMnBi$_2$ with the $I4/mmm$ lattice \cite{Rahn2017}. 

\begin{figure}
\includegraphics[width=7cm]{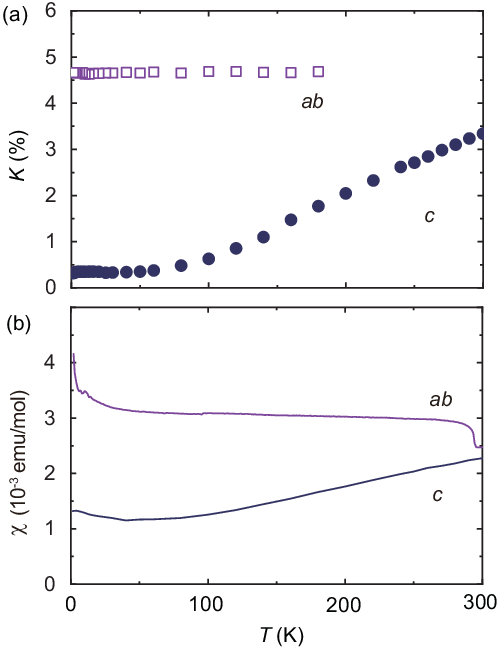}
 \caption{\label{FIG2}(a) Temperature $T$ dependence of $^{209}$Bi Knight shift $K$ defined as the relative frequency shift $K = (\omega - \omega_0)/\omega_0$ ($\omega_0 = \gamma_{\rm n} H$) in a magnetic field parallel or perpendicular to the $c$ axis in BaMnBi$_2$. (b) Magnetic susceptibility obtained from the bulk magnetization measurement for magnetic field along the $c$ axis or the $ab$ plane.}
\end{figure}

The Knight shift $K$ obtained from the central line of the $^{209}$Bi NMR spectrum is shown in Fig. \ref{FIG2}(a). $K$ reaches 3.3\% along the $c$ axis at 300 K and linearly decreases with temperature $T$. Then it becomes constant below 40 K. A similar behavior was observed in the magnetic susceptibility $\chi$ obtained from the magnetization [Fig. \ref{FIG2}(b)]. Since $\chi$ is dominated by Mn spins aligned along the $c$ axis, $K$ predominantly comes from the transferred hyperfine field from Mn spins, where the coupling constant was evaluated as $A_{\rm hf} = 14.2(5)$ T/$\mu_{\rm B}$ from the $K$-$\chi$ linearity \cite{SM}. In contrast, $K$ along the $ab$ plane maintains a high constant value (4.6\%) below 180 K due to the canting of Mn moments below $T_{\rm N}$ \cite{Kondo2021Bi, chen2017Bi, Liu2017, Rahn2017}. $K$ from the on-site Bi(1) spins should also be linear to $T$ but isotropic \cite{SM}, which is masked by the large contribution of Mn. 

\begin{figure*} 
\includegraphics[width=13cm]{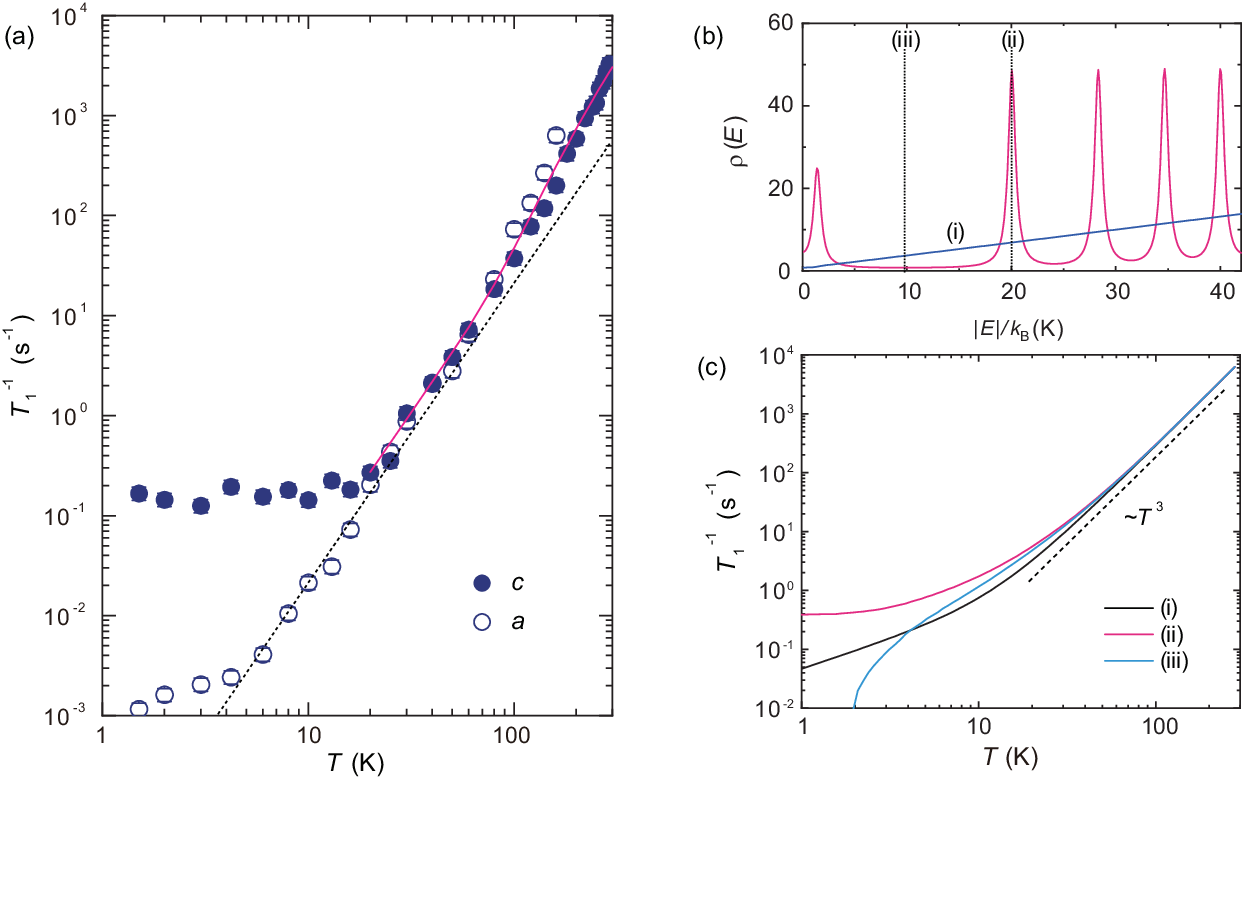}
\caption{\label{FIG3} (a) $^{209}$Bi nuclear spin-lattice relaxation rate $1/T_1$ measured under magnetic field (8.51 T) parallel (open circles) and normal (closed circles) to the $ab$ plane in BaMnBi$_2$. A dotted line represents a guide of $T^3$. A red curve represents a fitting by the gapped spin-wave excitation in addition to Dirac fermions. (b) Density of states $\rho(E)$ linear to $|E|$ in the absence of magnetic field (i), where the chemical potential $\mu$ is located slightly above (or below) the Dirac point. Quantized density of states following Eq.(2) with $\mu$ crossing the second Landau level (ii) and between Landau levels (iii). A small band gap $\Delta/k_{\rm B} = 1.4$ K is assumed. (c) Temperature dependence of $1/T_1$ calculated using Eq.(1) under the in-plane magnetic field without the Landau quantization (i, black), under magnetic field along $c$ axis with $\mu/k_{\rm B} = 20$ K (ii, red), and for $\mu/k_{\rm B} = 10$ K located in between the Landau levels (iii, blue). }
\end{figure*}

Low-energy excitation is investigated by $1/T_1$ in a magnetic field parallel or normal to the conducting layer, as shown in Fig. \ref{FIG3}. At high temperatures, $1/T_1$ exceeds $10^3$ s$^{-1}$. The extremely large $1/T_1$, despite a semimetal with a reduced density of states, can be attributed to significant antiferromagnetic fluctuations and magnon excitations around $T_{\rm N}$. Indeed, $1/T_1$ was fitted by an exponential function $\propto T^2{\rm exp}(-\Delta/k_{\rm B}T)$ for the two-magnon process with $\Delta = 37(3)$ meV \cite{Beeman1968} in addition to the $T^3$ dependence for Dirac fermions. The magnon contribution is rapidly suppressed below 100 K, $1/T_1$ is governed by the spin excitation of 2D Dirac fermions. Here the electric quadrupole relaxation would be negligible compared to the spin-flip process in the absence of the lattice softening \cite{Maebashi2019}. 

Below 100 K, $1/T_1$ is nearly isotropic and obeys a power law ($\propto T^3$), as expected in 2D Dirac semimetals with linear band dispersion \cite{dora2009graphene, Maebashi2019}. Notably, the $T^3$ dependence of $1/T_1$ persists to low temperatures ($\sim 6$ K) under the magnetic field along the $ab$ plane. Below 6 K, $1/T_1$ exhibits the Korringa law ($1/T_1 \sim T$), reflecting a residual density of states. Thus, the present system is regarded as an ideal Dirac semimetal with the Fermi level close to the Dirac point. 

In a magnetic field normal to the $ab$ plane, $1/T_1$ becomes constant below $20$ K. At the lowest temperature ($T = 1.5$ K), the anisotropy of $1/T_1$ exceeds $\sim 10^2$. The constant $1/T_1$ means an increase of $1/T_1T (\propto T^{-1}$) proportional to the square of the density of states. This behavior strikingly differs from conventional metal following $1/T_1T$ = constant and represents the squeezing of low-lying excitation into a discrete Landau level with energy damping $\Gamma < k_{\rm B}T$ where the SdH oscillation becomes sharpened and the Hall resistance shows plateaus \cite{li2016Bi, Kondo2021Bi}. 

\begin{figure*}
\includegraphics[width=13.5cm]{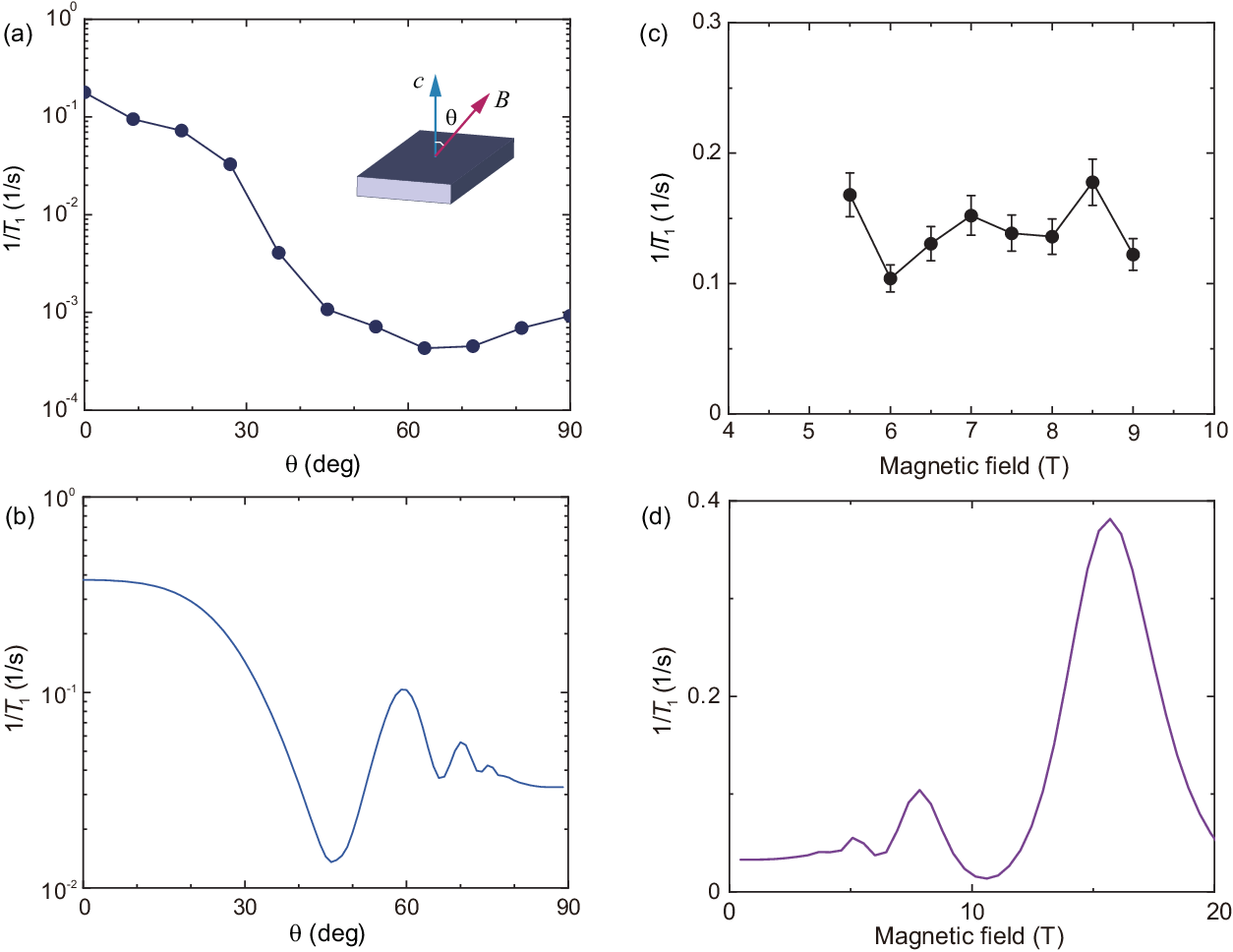}
\caption{\label{FIG4} Angular dependence of $1/T_1$ for BaMnBi$_2$ at $1.5$ K and $8.5$ T, where the magnetic field is tilted from the $c$ axis. (b) Numerical calculation of $1/T_1$, based on Eq.(\ref{DOS_Landau}) for quantized 2D Dirac fermions, as a function of the out-of-plane component of magnetic field for $v_{\rm F} = 1.6 \times 10^5$ m/s \cite{sharapov2004Dirac, li2016Bi}. (c) Magnetic field dependence of $1/T_1$ for BaMnBi$_2$ at $T=1.5$ K. (d) $1/T_1$ calculated for $\mu/k_{\rm B} =$ 20 K and the damping factor $\Gamma/k_{\rm B} = 0.2$ K.}
\end{figure*}

In 2D Dirac semimetals, $1/T_1$ measuring the dynamic spin susceptibility is given by \cite{dora2009graphene}
\begin{equation}
\label{T1}
\frac{1}{T_1}=\frac{A_{\rm hf}^2\pi k_BT}{\hbar}\int^\infty_{-\infty}\frac{\rho(E)^2 dE}{4k_BT{\rm cosh}^2[(E-\mu)/2k_BT]}.
\end{equation}
In the absence of Landau levels, the density of states $\rho(E)$ is linear to the energy $E$ around the Dirac point, as shown in Fig. \ref{FIG4}(a), and expressed as $\rho(E)=\frac{A_c|E|}{2\pi\hbar^2v_F^2}$ using the area of the unit cell $A_c$. Here, the Fermi velocity $v_F=1.6\times 10^5$ m/s was obtained from the resistivity measurement in BaMnBi$_2$ \cite{li2016Bi}. The calculated $1/T_1$ using Eq. (\ref{T1}) with a chemical potential $\mu$ = 20 K follows the $T^3$ dependence below 300 K and becomes linear for $T < 6$ K at zero field [Fig. \ref{FIG3}(c)], in good agreement with the experimental result for $B || ab$.  

The Dirac semimetals inherently involve low-energy excitations from spin and orbital fluctuations. The spin part follows $1/T_1 \sim T^3$ and $\sim T$ for $k_{\rm B}T  > \mu$ and $k_{\rm B}T < \mu$, respectively. The orbital part consists of the intraband transition in the presence of orbital degeneracy and the interband transition that produces huge diamagnetism \cite{dora2009graphene, Maebashi2019, hirosawa2017T1}. The latter is negligible due to orbital confinement in 2D systems, in contrast to 3D systems such as Bi$_{1-x}$Sb$_x$ with isotropic orbital fluctuations \cite{Watanabe}. 

Under a magnetic field normal to the $ab$ plane, the continuous energy distribution of $\rho(E)$ is quantized into Landau levels [Fig. \ref{FIG3}(b)]. Then, $\rho(E)$ in Eq. (1) is replaced by \cite{sharapov2004Dirac} 
\begin{eqnarray}
\label{DOS_Landau}
\rho(E) = \frac{A_c}{\pi^2}\left\{\Gamma {\rm ln}\frac{D^2}{2eH} 
-{\rm Im}\left[(\epsilon+i\Gamma)\left(\psi(x)  +\frac{2}{x}\right)\right]\right\}
\end{eqnarray}
using a digamma function $\psi(x)$ with $x = \left(\frac{\Delta^2-(\epsilon+i\Gamma)^2}{2eH}\right)$, a damping factor $\Gamma= 0.03E_F$ ($E_F$: Fermi energy), the energy cutoff $D = 1$ eV, and a band gap $\Delta/k_{\rm B} = 1.4$ K. $1/T_1$ shows a maximum when $\mu$ crosses the Landau level. The sharp spike $\rho(E)$ with $\Gamma < k_{\rm B}T$ gives the $T$-invariant $1/T_1$ similar to the behavior of localized moments, as indicated by a red curve in Fig. \ref{FIG3}(c). $1/T_1$ is suppressed for $\mu$ located between Landau levels at low temperatures. At 8.5 T, $\mu$ is located close to the second Landau level in reference to the SdH oscillation \cite{Kondo2021Bi}. 

We further investigate $1/T_1$ as a function of the field angle $\theta$ tilted from the $c$ axis as well as the field strength in Fig. \ref{FIG4}. For the highly 2D electronic structure of BaMnBi$_2$, the filling factor of the Landau level varies continuously by reducing the out-of-plane component of the magnetic field, $B{\rm cos}\theta$. As shown in Fig. \ref{FIG4}(a), $1/T_1$ is suppressed smoothly for $\theta < 20^\circ$ and decreases dramatically for $\theta > 30^\circ$. It exhibits a minimum around $\theta = 63^\circ$, where $1/T_1$ becomes 1/400 times smaller than the maximum value at $\theta = 0^\circ$, followed by an increase towards 90$^\circ$. 
 
The result is compared with a numerical calculation of $1/T_1$ using Eq. (\ref{DOS_Landau}) in Fig. \ref{FIG4}(b). By tilting the field direction from the $c$ axis, $\mu$ successively crosses higher Landau levels by reducing the out-of-plane field after exhibiting a minimum at 48$^\circ$, smaller than the experimental result. Since the minimum of $1/T_1$ corresponds to $\mu$ located between Landau levels, the larger $\mu$ lowers the minimum angle \cite{SM}. Thus, the Landau level at $\theta =0$ should be $n=1$, consistent with the main SdH oscillation \cite{Kondo2021Bi}. The difference between theoretical and experimental results is not simply explained by the sum of two Landau levels in different bands \cite{SM}, but by anisotropy of the Dirac cone and the spin splitting, as discussed in the following. 

 The magnetic field dependence of $1/T_1$ was measured in the field range of 5 - 9 T along the $c$ axis [Fig. \ref{FIG4}(c)], using the same pick-up coil to maintain the excitation condition. $1/T_1$ depends weakly on the magnetic field and increases at 5.5 and 8.5 T, consistent with the numerical calculation [Fig. \ref{FIG4}(d)]. Therefore, the $1/T_1$ enhancements occur at the Fermi level crossing the second and third Landau levels, in good agreement with the SdH oscillation in a low field range \cite{Kondo2021Bi}. 

The anisotropy of $1/T_1$ observed in BaMnBi$_2$ ($\simeq 150$) is much higher than the numerical calculation ($\simeq 11$) at 1.5 K, as shown in Figs. \ref{FIG3} and \ref{FIG4}. The difference suggests a significant role of spin-orbit coupling that enhances the Zeeman splitting. Since the spin splitting was not apparently resolved in the SdH oscillation \cite{li2016Bi,Kondo2021Bi}, the energy splitting should be comparable to the $\Gamma$ and $k_{\rm B}T$ scales. $1/T_1$ can be strongly enhanced by lifting the degeneracy via spin-orbit coupling \cite{Tsaran2014}, because of the high sensitivity to low-energy ($\omega \sim 0$) excitation.  

Another notable difference between theory and experiment is the onset temperature of $1/T_1$ = const. behavior. In the absence of spin-orbit coupling, the spin-split energy is equivalent to the Zeeman energy, and thus $1/T_1$ becomes constant for $k_{\rm B}T < g\mu_{\rm B}B$ much lower than $\mu$, as seen in Fig. \ref{FIG3}(b). Spin splitting can be largely enhanced in the presence of spin-orbit coupling [Fig. \ref{FIG1}(b)]. The excitation is analogous to a paramagnetic Mott insulator where the electron correlation induces the band (spin) splitting and $1/T_1$ = const. behavior, as observed below 20 K for $B || c$ in Fig. \ref{FIG3} (a). In the presence of the valley degrees of freedom, the valleys with different chirality have reversed spin-flip processes for spin-split Landau levels. 

Our result demonstrates that NMR spectroscopy serves as a sensitive probe for bulk spin excitation in the anisotropic spin-split Landau level. Only a few examples are known for quantized spin excitation by NMR \cite{Bridges1969, Berg1990, Fujii2023}, in which $1/T_1$ is enhanced by less than a few tens at the Landau level. Our result far exceeds previous reports because of the strong spin-orbit coupling and the well-separated discrete Landau level in the QHE regime. Thus, the present system provides the unique platform for investigating the effect of strong spin-orbit coupling in 2D Dirac fermions. A further experiment in higher magnetic fields is required to access the fully spin-polarized lowest Landau level. The present method will be applied to extensive topological materials including anomalous quantum (spin) Hall effect and quantum spin liquid. 

In conclusion, we investigated local spin susceptibility by $^{209}$Bi NMR measurements in the magnetic Dirac semimetal BaMnBi$_2$. It is uncovered that Dirac fermions in the Bi layer experience no staggered local fields from antiferromagnetic moments related by the mirror symmetry. The power-law dependence of $1/T_1$ in the extensive temperature range highlights the realization of the ideal two-dimensional Dirac fermion system. The spin excitation is enhanced by two orders of magnitude by Landau quantization at low temperatures. An indication of the quantum oscillation is observed in the field dependence of $1/T_1$. The results open experimental research for magnetism of the quantum Hall system in bulk topological materials.

The authors thank H. Sakai, A. Yamakage, Y. Fuseya, and A. Kobayashi for useful discussions. We acknowledge the support from the grant-in-aid in scientific research by JSPS (No.JP19H05824, No.23H04025, and No.24H00954). 

\bibliography{BaMnBi2}% Produces the bibliography via BibTeX.

\providecommand{\noopsort}[1]{}\providecommand{\singleletter}[1]{#1}%
\begin{thebibliography}{62}%
\makeatletter
\providecommand \@ifxundefined [1]{%
 \@ifx{#1\undefined}
}%
\providecommand \@ifnum [1]{%
 \ifnum #1\expandafter \@firstoftwo
 \else \expandafter \@secondoftwo
 \fi
}%
\providecommand \@ifx [1]{%
 \ifx #1\expandafter \@firstoftwo
 \else \expandafter \@secondoftwo
 \fi
}%
\providecommand \natexlab [1]{#1}%
\providecommand \enquote  [1]{``#1''}%
\providecommand \bibnamefont  [1]{#1}%
\providecommand \bibfnamefont [1]{#1}%
\providecommand \citenamefont [1]{#1}%
\providecommand \href@noop [0]{\@secondoftwo}%
\providecommand \href [0]{\begingroup \@sanitize@url \@href}%
\providecommand \@href[1]{\@@startlink{#1}\@@href}%
\providecommand \@@href[1]{\endgroup#1\@@endlink}%
\providecommand \@sanitize@url [0]{\catcode `\\12\catcode `\$12\catcode
  `\&12\catcode `\#12\catcode `\^12\catcode `\_12\catcode `\%12\relax}%
\providecommand \@@startlink[1]{}%
\providecommand \@@endlink[0]{}%
\providecommand \url  [0]{\begingroup\@sanitize@url \@url }%
\providecommand \@url [1]{\endgroup\@href {#1}{\urlprefix }}%
\providecommand \urlprefix  [0]{URL }%
\providecommand \Eprint [0]{\href }%
\providecommand \doibase [0]{https://doi.org/}%
\providecommand \selectlanguage [0]{\@gobble}%
\providecommand \bibinfo  [0]{\@secondoftwo}%
\providecommand \bibfield  [0]{\@secondoftwo}%
\providecommand \translation [1]{[#1]}%
\providecommand \BibitemOpen [0]{}%
\providecommand \bibitemStop [0]{}%
\providecommand \bibitemNoStop [0]{.\EOS\space}%
\providecommand \EOS [0]{\spacefactor3000\relax}%
\providecommand \BibitemShut  [1]{\csname bibitem#1\endcsname}%
\let\auto@bib@innerbib\@empty
%</preamble>
\bibitem [{\citenamefont {Klitzing}\ \emph {et~al.}(1980)\citenamefont
  {Klitzing}, \citenamefont {Dorda},\ and\ \citenamefont
  {Pepper}}]{vKlitzing1980}%
  \BibitemOpen
  \bibfield  {author} {\bibinfo {author} {\bibfnamefont {K.~v.}\ \bibnamefont
  {Klitzing}}, \bibinfo {author} {\bibfnamefont {G.}~\bibnamefont {Dorda}},\
  and\ \bibinfo {author} {\bibfnamefont {M.}~\bibnamefont {Pepper}},\
  }\bibfield  {title} {\bibinfo {title} {New method for high-accuracy
  determination of the fine-structure constant based on quantized {H}all
  resistance},\ }\href {https://doi.org/10.1103/PhysRevLett.45.494} {\bibfield
  {journal} {\bibinfo  {journal} {Phys. Rev. Lett.}\ }\textbf {\bibinfo
  {volume} {45}},\ \bibinfo {pages} {494} (\bibinfo {year} {1980})}\BibitemShut
  {NoStop}%
\bibitem [{\citenamefont {Tsui}\ \emph {et~al.}(1982)\citenamefont {Tsui},
  \citenamefont {Stormer},\ and\ \citenamefont {Gossard}}]{Tsui1982}%
  \BibitemOpen
  \bibfield  {author} {\bibinfo {author} {\bibfnamefont {D.~C.}\ \bibnamefont
  {Tsui}}, \bibinfo {author} {\bibfnamefont {H.~L.}\ \bibnamefont {Stormer}},\
  and\ \bibinfo {author} {\bibfnamefont {A.~C.}\ \bibnamefont {Gossard}},\
  }\bibfield  {title} {\bibinfo {title} {Two-dimensional magnetotransport in
  the extreme quantum limit},\ }\href
  {https://doi.org/10.1103/PhysRevLett.48.1559} {\bibfield  {journal} {\bibinfo
   {journal} {Phys. Rev. Lett.}\ }\textbf {\bibinfo {volume} {48}},\ \bibinfo
  {pages} {1559} (\bibinfo {year} {1982})}\BibitemShut {NoStop}%
\bibitem [{\citenamefont {von Klitzing}\ \emph {et~al.}(2020)\citenamefont {von
  Klitzing}, \citenamefont {Chakraborty}, \citenamefont {Kim}, \citenamefont
  {Madhavan}, \citenamefont {Dai}, \citenamefont {McIver}, \citenamefont
  {Tokura}, \citenamefont {Savary}, \citenamefont {Smirnova}, \citenamefont
  {Rey}, \citenamefont {Felser}, \citenamefont {Gooth},\ and\ \citenamefont
  {Qi}}]{vonKlitzing2020}%
  \BibitemOpen
  \bibfield  {author} {\bibinfo {author} {\bibfnamefont {K.}~\bibnamefont {von
  Klitzing}}, \bibinfo {author} {\bibfnamefont {T.}~\bibnamefont
  {Chakraborty}}, \bibinfo {author} {\bibfnamefont {P.}~\bibnamefont {Kim}},
  \bibinfo {author} {\bibfnamefont {V.}~\bibnamefont {Madhavan}}, \bibinfo
  {author} {\bibfnamefont {X.}~\bibnamefont {Dai}}, \bibinfo {author}
  {\bibfnamefont {J.}~\bibnamefont {McIver}}, \bibinfo {author} {\bibfnamefont
  {Y.}~\bibnamefont {Tokura}}, \bibinfo {author} {\bibfnamefont
  {L.}~\bibnamefont {Savary}}, \bibinfo {author} {\bibfnamefont
  {D.}~\bibnamefont {Smirnova}}, \bibinfo {author} {\bibfnamefont {A.~M.}\
  \bibnamefont {Rey}}, \bibinfo {author} {\bibfnamefont {C.}~\bibnamefont
  {Felser}}, \bibinfo {author} {\bibfnamefont {J.}~\bibnamefont {Gooth}},\ and\
  \bibinfo {author} {\bibfnamefont {X.}~\bibnamefont {Qi}},\ }\bibfield
  {title} {\bibinfo {title} {40 years of the quantum {H}all effect},\ }\href
  {https://doi.org/10.1038/s42254-020-0209-1} {\bibfield  {journal} {\bibinfo
  {journal} {Nat. Rev. Phys.}\ }\textbf {\bibinfo {volume} {2}},\ \bibinfo
  {pages} {397} (\bibinfo {year} {2020})}\BibitemShut {NoStop}%
\bibitem [{\citenamefont {Barrett}\ \emph {et~al.}(1996)\citenamefont
  {Barrett}, \citenamefont {Dabbagh}, \citenamefont {Pfeiffer}, \citenamefont
  {West},\ and\ \citenamefont {Tycko}}]{Barrett1996}%
  \BibitemOpen
  \bibfield  {author} {\bibinfo {author} {\bibfnamefont {S.}~\bibnamefont
  {Barrett}}, \bibinfo {author} {\bibfnamefont {G.}~\bibnamefont {Dabbagh}},
  \bibinfo {author} {\bibfnamefont {L.}~\bibnamefont {Pfeiffer}}, \bibinfo
  {author} {\bibfnamefont {K.}~\bibnamefont {West}},\ and\ \bibinfo {author}
  {\bibfnamefont {R.}~\bibnamefont {Tycko}},\ }\bibfield  {title} {\bibinfo
  {title} {{NMR} measurement of the spin magnetization and spin dynamics in the
  quantum {H}all regimes},\ }\href
  {https://doi.org/https://doi.org/10.1016/0039-6028(96)00398-6} {\bibfield
  {journal} {\bibinfo  {journal} {Sur. Sci.}\ }\textbf {\bibinfo {volume}
  {361-362}},\ \bibinfo {pages} {261} (\bibinfo {year} {1996})}\BibitemShut
  {NoStop}%
\bibitem [{\citenamefont {Kronm\"uller}\ \emph {et~al.}(1999)\citenamefont
  {Kronm\"uller}, \citenamefont {Dietsche}, \citenamefont {v.~Klitzing},
  \citenamefont {Denninger}, \citenamefont {Wegscheider},\ and\ \citenamefont
  {Bichler}}]{Kronmuller1999}%
  \BibitemOpen
  \bibfield  {author} {\bibinfo {author} {\bibfnamefont {S.}~\bibnamefont
  {Kronm\"uller}}, \bibinfo {author} {\bibfnamefont {W.}~\bibnamefont
  {Dietsche}}, \bibinfo {author} {\bibfnamefont {K.}~\bibnamefont
  {v.~Klitzing}}, \bibinfo {author} {\bibfnamefont {G.}~\bibnamefont
  {Denninger}}, \bibinfo {author} {\bibfnamefont {W.}~\bibnamefont
  {Wegscheider}},\ and\ \bibinfo {author} {\bibfnamefont {M.}~\bibnamefont
  {Bichler}},\ }\bibfield  {title} {\bibinfo {title} {New type of electron
  nuclear-spin interaction from resistively detected {NMR} in the fractional
  quantum {H}all effect regime},\ }\href
  {https://doi.org/10.1103/PhysRevLett.82.4070} {\bibfield  {journal} {\bibinfo
   {journal} {Phys. Rev. Lett.}\ }\textbf {\bibinfo {volume} {82}},\ \bibinfo
  {pages} {4070} (\bibinfo {year} {1999})}\BibitemShut {NoStop}%
\bibitem [{\citenamefont {Hashimoto}\ \emph {et~al.}(2002)\citenamefont
  {Hashimoto}, \citenamefont {Muraki}, \citenamefont {Saku},\ and\
  \citenamefont {Hirayama}}]{Hashimoto2002NMR}%
  \BibitemOpen
  \bibfield  {author} {\bibinfo {author} {\bibfnamefont {K.}~\bibnamefont
  {Hashimoto}}, \bibinfo {author} {\bibfnamefont {K.}~\bibnamefont {Muraki}},
  \bibinfo {author} {\bibfnamefont {T.}~\bibnamefont {Saku}},\ and\ \bibinfo
  {author} {\bibfnamefont {Y.}~\bibnamefont {Hirayama}},\ }\bibfield  {title}
  {\bibinfo {title} {Electrically controlled nuclear spin polarization and
  relaxation by quantum-{H}all states},\ }\href
  {https://doi.org/10.1103/PhysRevLett.88.176601} {\bibfield  {journal}
  {\bibinfo  {journal} {Phys. Rev. Lett.}\ }\textbf {\bibinfo {volume} {88}},\
  \bibinfo {pages} {176601} (\bibinfo {year} {2002})}\BibitemShut {NoStop}%
\bibitem [{\citenamefont {Li}\ and\ \citenamefont
  {Andrei}(2007)}]{li2007graphite}%
  \BibitemOpen
  \bibfield  {author} {\bibinfo {author} {\bibfnamefont {G.}~\bibnamefont
  {Li}}\ and\ \bibinfo {author} {\bibfnamefont {E.~Y.}\ \bibnamefont
  {Andrei}},\ }\bibfield  {title} {\bibinfo {title} {Observation of {L}andau
  levels of {D}irac fermions in graphite},\ }\href@noop {} {\bibfield
  {journal} {\bibinfo  {journal} {Nat. Phys.}\ }\textbf {\bibinfo {volume}
  {3}},\ \bibinfo {pages} {623} (\bibinfo {year} {2007})}\BibitemShut {NoStop}%
\bibitem [{\citenamefont {Melcer}\ \emph {et~al.}(2024)\citenamefont {Melcer},
  \citenamefont {Gil}, \citenamefont {Paul}, \citenamefont {Tiwari},
  \citenamefont {Umansky}, \citenamefont {Heiblum}, \citenamefont {Oreg},
  \citenamefont {Stern},\ and\ \citenamefont {Berg}}]{Melcer2024}%
  \BibitemOpen
  \bibfield  {author} {\bibinfo {author} {\bibfnamefont {R.~A.}\ \bibnamefont
  {Melcer}}, \bibinfo {author} {\bibfnamefont {A.}~\bibnamefont {Gil}},
  \bibinfo {author} {\bibfnamefont {A.~K.}\ \bibnamefont {Paul}}, \bibinfo
  {author} {\bibfnamefont {P.}~\bibnamefont {Tiwari}}, \bibinfo {author}
  {\bibfnamefont {V.}~\bibnamefont {Umansky}}, \bibinfo {author} {\bibfnamefont
  {M.}~\bibnamefont {Heiblum}}, \bibinfo {author} {\bibfnamefont
  {Y.}~\bibnamefont {Oreg}}, \bibinfo {author} {\bibfnamefont {A.}~\bibnamefont
  {Stern}},\ and\ \bibinfo {author} {\bibfnamefont {E.}~\bibnamefont {Berg}},\
  }\bibfield  {title} {\bibinfo {title} {Heat conductance of the quantum {H}all
  bulk},\ }\href {https://doi.org/10.1038/s41586-023-06858-z} {\bibfield
  {journal} {\bibinfo  {journal} {Nature}\ }\textbf {\bibinfo {volume} {625}},\
  \bibinfo {pages} {489} (\bibinfo {year} {2024})}\BibitemShut {NoStop}%
\bibitem [{\citenamefont {Moore}\ and\ \citenamefont {Read}(1991)}]{Moore1991}%
  \BibitemOpen
  \bibfield  {author} {\bibinfo {author} {\bibfnamefont {G.}~\bibnamefont
  {Moore}}\ and\ \bibinfo {author} {\bibfnamefont {N.}~\bibnamefont {Read}},\
  }\bibfield  {title} {\bibinfo {title} {Nonabelions in the fractional quantum
  {H}all effect},\ }\href
  {https://doi.org/https://doi.org/10.1016/0550-3213(91)90407-O} {\bibfield
  {journal} {\bibinfo  {journal} {Nucl. Phys. B}\ }\textbf {\bibinfo {volume}
  {360}},\ \bibinfo {pages} {362} (\bibinfo {year} {1991})}\BibitemShut
  {NoStop}%
\bibitem [{\citenamefont {Wen}(1995)}]{Wen1995}%
  \BibitemOpen
  \bibfield  {author} {\bibinfo {author} {\bibfnamefont {X.-G.}\ \bibnamefont
  {Wen}},\ }\bibfield  {title} {\bibinfo {title} {Topological orders and edge
  excitations in fractional quantum {H}all states},\ }\href
  {https://doi.org/10.1080/00018739500101566} {\bibfield  {journal} {\bibinfo
  {journal} {Adv. Phys.}\ }\textbf {\bibinfo {volume} {44}},\ \bibinfo {pages}
  {405} (\bibinfo {year} {1995})}\BibitemShut {NoStop}%
\bibitem [{\citenamefont {Novoselov}\ \emph {et~al.}(2007)\citenamefont
  {Novoselov}, \citenamefont {Jiang}, \citenamefont {Zhang}, \citenamefont
  {Morozov}, \citenamefont {Stormer}, \citenamefont {Zeitler}, \citenamefont
  {Maan}, \citenamefont {Boebinger}, \citenamefont {Kim},\ and\ \citenamefont
  {Geim}}]{Novoselov2007}%
  \BibitemOpen
  \bibfield  {author} {\bibinfo {author} {\bibfnamefont {K.~S.}\ \bibnamefont
  {Novoselov}}, \bibinfo {author} {\bibfnamefont {Z.}~\bibnamefont {Jiang}},
  \bibinfo {author} {\bibfnamefont {Y.}~\bibnamefont {Zhang}}, \bibinfo
  {author} {\bibfnamefont {S.~V.}\ \bibnamefont {Morozov}}, \bibinfo {author}
  {\bibfnamefont {H.~L.}\ \bibnamefont {Stormer}}, \bibinfo {author}
  {\bibfnamefont {U.}~\bibnamefont {Zeitler}}, \bibinfo {author} {\bibfnamefont
  {J.~C.}\ \bibnamefont {Maan}}, \bibinfo {author} {\bibfnamefont {G.~S.}\
  \bibnamefont {Boebinger}}, \bibinfo {author} {\bibfnamefont {P.}~\bibnamefont
  {Kim}},\ and\ \bibinfo {author} {\bibfnamefont {A.~K.}\ \bibnamefont
  {Geim}},\ }\bibfield  {title} {\bibinfo {title} {Room-temperature quantum
  {H}all effect in graphene},\ }\href@noop {} {\bibfield  {journal} {\bibinfo
  {journal} {Science}\ }\textbf {\bibinfo {volume} {315}},\ \bibinfo {pages}
  {1379} (\bibinfo {year} {2007})}\BibitemShut {NoStop}%
\bibitem [{\citenamefont {Bolotin}\ \emph {et~al.}(2009)\citenamefont
  {Bolotin}, \citenamefont {Ghahari}, \citenamefont {Shulman}, \citenamefont
  {Stormer},\ and\ \citenamefont {Kim}}]{Bolotin2009}%
  \BibitemOpen
  \bibfield  {author} {\bibinfo {author} {\bibfnamefont {K.~I.}\ \bibnamefont
  {Bolotin}}, \bibinfo {author} {\bibfnamefont {F.}~\bibnamefont {Ghahari}},
  \bibinfo {author} {\bibfnamefont {M.~D.}\ \bibnamefont {Shulman}}, \bibinfo
  {author} {\bibfnamefont {H.~L.}\ \bibnamefont {Stormer}},\ and\ \bibinfo
  {author} {\bibfnamefont {P.}~\bibnamefont {Kim}},\ }\bibfield  {title}
  {\bibinfo {title} {Observation of the fractional quantum {H}all effect in
  graphene},\ }\href {https://doi.org/10.1038/nature08582} {\bibfield
  {journal} {\bibinfo  {journal} {Nature}\ }\textbf {\bibinfo {volume} {462}},\
  \bibinfo {pages} {196} (\bibinfo {year} {2009})}\BibitemShut {NoStop}%
\bibitem [{\citenamefont {Castro~Neto}\ \emph {et~al.}(2009)\citenamefont
  {Castro~Neto}, \citenamefont {Guinea}, \citenamefont {Peres}, \citenamefont
  {Novoselov},\ and\ \citenamefont {Geim}}]{Castro2009}%
  \BibitemOpen
  \bibfield  {author} {\bibinfo {author} {\bibfnamefont {A.~H.}\ \bibnamefont
  {Castro~Neto}}, \bibinfo {author} {\bibfnamefont {F.}~\bibnamefont {Guinea}},
  \bibinfo {author} {\bibfnamefont {N.~M.~R.}\ \bibnamefont {Peres}}, \bibinfo
  {author} {\bibfnamefont {K.~S.}\ \bibnamefont {Novoselov}},\ and\ \bibinfo
  {author} {\bibfnamefont {A.~K.}\ \bibnamefont {Geim}},\ }\bibfield  {title}
  {\bibinfo {title} {The electronic properties of graphene},\ }\href
  {https://doi.org/10.1103/RevModPhys.81.109} {\bibfield  {journal} {\bibinfo
  {journal} {Rev. Mod. Phys.}\ }\textbf {\bibinfo {volume} {81}},\ \bibinfo
  {pages} {109} (\bibinfo {year} {2009})}\BibitemShut {NoStop}%
\bibitem [{\citenamefont {He}\ \emph {et~al.}(2014)\citenamefont {He},
  \citenamefont {Hong}, \citenamefont {Dong}, \citenamefont {Pan},
  \citenamefont {Zhang}, \citenamefont {Zhang},\ and\ \citenamefont
  {Li}}]{He2014}%
  \BibitemOpen
  \bibfield  {author} {\bibinfo {author} {\bibfnamefont {L.~P.}\ \bibnamefont
  {He}}, \bibinfo {author} {\bibfnamefont {X.~C.}\ \bibnamefont {Hong}},
  \bibinfo {author} {\bibfnamefont {J.~K.}\ \bibnamefont {Dong}}, \bibinfo
  {author} {\bibfnamefont {J.}~\bibnamefont {Pan}}, \bibinfo {author}
  {\bibfnamefont {Z.}~\bibnamefont {Zhang}}, \bibinfo {author} {\bibfnamefont
  {J.}~\bibnamefont {Zhang}},\ and\ \bibinfo {author} {\bibfnamefont {S.~Y.}\
  \bibnamefont {Li}},\ }\bibfield  {title} {\bibinfo {title} {Quantum transport
  evidence for the three-dimensional {D}irac semimetal phase in
  {C}d$_{3}${A}s$_{2}$},\ }\href
  {https://doi.org/10.1103/PhysRevLett.113.246402} {\bibfield  {journal}
  {\bibinfo  {journal} {Phys. Rev. Lett.}\ }\textbf {\bibinfo {volume} {113}},\
  \bibinfo {pages} {246402} (\bibinfo {year} {2014})}\BibitemShut {NoStop}%
\bibitem [{\citenamefont {Jeon}\ \emph {et~al.}(2014)\citenamefont {Jeon},
  \citenamefont {Zhou}, \citenamefont {Gyenis}, \citenamefont {Feldman},
  \citenamefont {Kimchi}, \citenamefont {Potter}, \citenamefont {Gibson},
  \citenamefont {Cava}, \citenamefont {Vishwanath},\ and\ \citenamefont
  {Yazdani}}]{Jeon2014}%
  \BibitemOpen
  \bibfield  {author} {\bibinfo {author} {\bibfnamefont {S.}~\bibnamefont
  {Jeon}}, \bibinfo {author} {\bibfnamefont {B.~B.}\ \bibnamefont {Zhou}},
  \bibinfo {author} {\bibfnamefont {A.}~\bibnamefont {Gyenis}}, \bibinfo
  {author} {\bibfnamefont {B.~E.}\ \bibnamefont {Feldman}}, \bibinfo {author}
  {\bibfnamefont {I.}~\bibnamefont {Kimchi}}, \bibinfo {author} {\bibfnamefont
  {A.~C.}\ \bibnamefont {Potter}}, \bibinfo {author} {\bibfnamefont {Q.~D.}\
  \bibnamefont {Gibson}}, \bibinfo {author} {\bibfnamefont {R.~J.}\
  \bibnamefont {Cava}}, \bibinfo {author} {\bibfnamefont {A.}~\bibnamefont
  {Vishwanath}},\ and\ \bibinfo {author} {\bibfnamefont {A.}~\bibnamefont
  {Yazdani}},\ }\bibfield  {title} {\bibinfo {title} {Landau quantization and
  quasiparticle interference in the three-dimensional {D}irac semimetal
  {C}d$_3${A}s$_2$},\ }\href {https://doi.org/10.1038/nmat4023} {\bibfield
  {journal} {\bibinfo  {journal} {Nat. Mater.}\ }\textbf {\bibinfo {volume}
  {13}},\ \bibinfo {pages} {851} (\bibinfo {year} {2014})}\BibitemShut
  {NoStop}%
\bibitem [{\citenamefont {Li}\ \emph {et~al.}(2015)\citenamefont {Li},
  \citenamefont {Chen}, \citenamefont {Meng}, \citenamefont {Guo},
  \citenamefont {Huang}, \citenamefont {Liu}, \citenamefont {Wang},\ and\
  \citenamefont {Chen}}]{li2015graphene}%
  \BibitemOpen
  \bibfield  {author} {\bibinfo {author} {\bibfnamefont {Z.}~\bibnamefont
  {Li}}, \bibinfo {author} {\bibfnamefont {L.}~\bibnamefont {Chen}}, \bibinfo
  {author} {\bibfnamefont {S.}~\bibnamefont {Meng}}, \bibinfo {author}
  {\bibfnamefont {L.}~\bibnamefont {Guo}}, \bibinfo {author} {\bibfnamefont
  {J.}~\bibnamefont {Huang}}, \bibinfo {author} {\bibfnamefont
  {Y.}~\bibnamefont {Liu}}, \bibinfo {author} {\bibfnamefont {W.}~\bibnamefont
  {Wang}},\ and\ \bibinfo {author} {\bibfnamefont {X.}~\bibnamefont {Chen}},\
  }\bibfield  {title} {\bibinfo {title} {Field and temperature dependence of
  intrinsic diamagnetism in graphene: Theory and experiment},\ }\href@noop {}
  {\bibfield  {journal} {\bibinfo  {journal} {Phys. Rev. B}\ }\textbf {\bibinfo
  {volume} {91}},\ \bibinfo {pages} {094429} (\bibinfo {year}
  {2015})}\BibitemShut {NoStop}%
\bibitem [{\citenamefont {Uchida}\ \emph {et~al.}(2017)\citenamefont {Uchida},
  \citenamefont {Nakazawa}, \citenamefont {Nishihaya}, \citenamefont {Akiba},
  \citenamefont {Kriener}, \citenamefont {Kozuka}, \citenamefont {Miyake},
  \citenamefont {Taguchi}, \citenamefont {Tokunaga}, \citenamefont {Nagaosa},
  \citenamefont {Tokura},\ and\ \citenamefont {Kawasaki}}]{Uchida2017}%
  \BibitemOpen
  \bibfield  {author} {\bibinfo {author} {\bibfnamefont {M.}~\bibnamefont
  {Uchida}}, \bibinfo {author} {\bibfnamefont {Y.}~\bibnamefont {Nakazawa}},
  \bibinfo {author} {\bibfnamefont {S.}~\bibnamefont {Nishihaya}}, \bibinfo
  {author} {\bibfnamefont {K.}~\bibnamefont {Akiba}}, \bibinfo {author}
  {\bibfnamefont {M.}~\bibnamefont {Kriener}}, \bibinfo {author} {\bibfnamefont
  {Y.}~\bibnamefont {Kozuka}}, \bibinfo {author} {\bibfnamefont
  {A.}~\bibnamefont {Miyake}}, \bibinfo {author} {\bibfnamefont
  {Y.}~\bibnamefont {Taguchi}}, \bibinfo {author} {\bibfnamefont
  {M.}~\bibnamefont {Tokunaga}}, \bibinfo {author} {\bibfnamefont
  {N.}~\bibnamefont {Nagaosa}}, \bibinfo {author} {\bibfnamefont
  {Y.}~\bibnamefont {Tokura}},\ and\ \bibinfo {author} {\bibfnamefont
  {M.}~\bibnamefont {Kawasaki}},\ }\bibfield  {title} {\bibinfo {title}
  {Quantum {H}all states observed in thin films of {D}irac semimetal
  {C}d$_3${A}s$_2$},\ }\href {https://doi.org/10.1038/s41467-017-02423-1}
  {\bibfield  {journal} {\bibinfo  {journal} {Nat. Commun.}\ }\textbf {\bibinfo
  {volume} {8}},\ \bibinfo {pages} {2274} (\bibinfo {year} {2017})}\BibitemShut
  {NoStop}%
\bibitem [{\citenamefont {Kealhofer}\ \emph {et~al.}(2020)\citenamefont
  {Kealhofer}, \citenamefont {Galletti}, \citenamefont {Schumann},
  \citenamefont {Suslov},\ and\ \citenamefont
  {Stemmer}}]{Kealhofer2020topological}%
  \BibitemOpen
  \bibfield  {author} {\bibinfo {author} {\bibfnamefont {D.~A.}\ \bibnamefont
  {Kealhofer}}, \bibinfo {author} {\bibfnamefont {L.}~\bibnamefont {Galletti}},
  \bibinfo {author} {\bibfnamefont {T.}~\bibnamefont {Schumann}}, \bibinfo
  {author} {\bibfnamefont {A.}~\bibnamefont {Suslov}},\ and\ \bibinfo {author}
  {\bibfnamefont {S.}~\bibnamefont {Stemmer}},\ }\bibfield  {title} {\bibinfo
  {title} {Topological insulator state and collapse of the quantum {H}all
  effect in a three-dimensional {D}irac semimetal heterojunction},\ }\href
  {https://doi.org/10.1103/PhysRevX.10.011050} {\bibfield  {journal} {\bibinfo
  {journal} {Phys. Rev. X}\ }\textbf {\bibinfo {volume} {10}},\ \bibinfo
  {pages} {011050} (\bibinfo {year} {2020})}\BibitemShut {NoStop}%
\bibitem [{\citenamefont {Liu}\ \emph {et~al.}(2022)\citenamefont {Liu},
  \citenamefont {Farahi}, \citenamefont {Chiu}, \citenamefont {Papic},
  \citenamefont {Watanabe}, \citenamefont {Taniguchi}, \citenamefont
  {Zaletel},\ and\ \citenamefont {Yazdani}}]{Liu2022}%
  \BibitemOpen
  \bibfield  {author} {\bibinfo {author} {\bibfnamefont {X.}~\bibnamefont
  {Liu}}, \bibinfo {author} {\bibfnamefont {G.}~\bibnamefont {Farahi}},
  \bibinfo {author} {\bibfnamefont {C.-L.}\ \bibnamefont {Chiu}}, \bibinfo
  {author} {\bibfnamefont {Z.}~\bibnamefont {Papic}}, \bibinfo {author}
  {\bibfnamefont {K.}~\bibnamefont {Watanabe}}, \bibinfo {author}
  {\bibfnamefont {T.}~\bibnamefont {Taniguchi}}, \bibinfo {author}
  {\bibfnamefont {M.~P.}\ \bibnamefont {Zaletel}},\ and\ \bibinfo {author}
  {\bibfnamefont {A.}~\bibnamefont {Yazdani}},\ }\bibfield  {title} {\bibinfo
  {title} {Visualizing broken symmetry and topological defects in a quantum
  {H}all ferromagnet},\ }\href {https://doi.org/10.1126/science.abm3770}
  {\bibfield  {journal} {\bibinfo  {journal} {Science}\ }\textbf {\bibinfo
  {volume} {375}},\ \bibinfo {pages} {321} (\bibinfo {year}
  {2022})}\BibitemShut {NoStop}%
\bibitem [{\citenamefont {Young}\ and\ \citenamefont {Kane}(2015)}]{Young2015}%
  \BibitemOpen
  \bibfield  {author} {\bibinfo {author} {\bibfnamefont {S.~M.}\ \bibnamefont
  {Young}}\ and\ \bibinfo {author} {\bibfnamefont {C.~L.}\ \bibnamefont
  {Kane}},\ }\bibfield  {title} {\bibinfo {title} {Dirac semimetals in two
  dimensions},\ }\href {https://doi.org/10.1103/PhysRevLett.115.126803}
  {\bibfield  {journal} {\bibinfo  {journal} {Phys. Rev. Lett.}\ }\textbf
  {\bibinfo {volume} {115}},\ \bibinfo {pages} {126803} (\bibinfo {year}
  {2015})}\BibitemShut {NoStop}%
\bibitem [{\citenamefont {Tajima}\ \emph {et~al.}(2013)\citenamefont {Tajima},
  \citenamefont {Yamauchi}, \citenamefont {Yamaguchi}, \citenamefont {Suda},
  \citenamefont {Kawasugi}, \citenamefont {Yamamoto}, \citenamefont {Kato},
  \citenamefont {Nishio},\ and\ \citenamefont {Kajita}}]{Tajima2013}%
  \BibitemOpen
  \bibfield  {author} {\bibinfo {author} {\bibfnamefont {N.}~\bibnamefont
  {Tajima}}, \bibinfo {author} {\bibfnamefont {T.}~\bibnamefont {Yamauchi}},
  \bibinfo {author} {\bibfnamefont {T.}~\bibnamefont {Yamaguchi}}, \bibinfo
  {author} {\bibfnamefont {M.}~\bibnamefont {Suda}}, \bibinfo {author}
  {\bibfnamefont {Y.}~\bibnamefont {Kawasugi}}, \bibinfo {author}
  {\bibfnamefont {H.~M.}\ \bibnamefont {Yamamoto}}, \bibinfo {author}
  {\bibfnamefont {R.}~\bibnamefont {Kato}}, \bibinfo {author} {\bibfnamefont
  {Y.}~\bibnamefont {Nishio}},\ and\ \bibinfo {author} {\bibfnamefont
  {K.}~\bibnamefont {Kajita}},\ }\bibfield  {title} {\bibinfo {title} {Quantum
  {H}all effect in multilayered massless {D}irac fermion systems with tilted
  cones},\ }\href {https://doi.org/10.1103/PhysRevB.88.075315} {\bibfield
  {journal} {\bibinfo  {journal} {Phys. Rev. B}\ }\textbf {\bibinfo {volume}
  {88}},\ \bibinfo {pages} {075315} (\bibinfo {year} {2013})}\BibitemShut
  {NoStop}%
\bibitem [{\citenamefont {Masuda}\ \emph {et~al.}(2016)\citenamefont {Masuda},
  \citenamefont {Sakai}, \citenamefont {Tokunaga}, \citenamefont {Yamasaki},
  \citenamefont {Miyake}, \citenamefont {Shiogai}, \citenamefont {Nakamura},
  \citenamefont {Awaji}, \citenamefont {Tsukazaki}, \citenamefont {Nakao} \emph
  {et~al.}}]{masuda2016QHE}%
  \BibitemOpen
  \bibfield  {author} {\bibinfo {author} {\bibfnamefont {H.}~\bibnamefont
  {Masuda}}, \bibinfo {author} {\bibfnamefont {H.}~\bibnamefont {Sakai}},
  \bibinfo {author} {\bibfnamefont {M.}~\bibnamefont {Tokunaga}}, \bibinfo
  {author} {\bibfnamefont {Y.}~\bibnamefont {Yamasaki}}, \bibinfo {author}
  {\bibfnamefont {A.}~\bibnamefont {Miyake}}, \bibinfo {author} {\bibfnamefont
  {J.}~\bibnamefont {Shiogai}}, \bibinfo {author} {\bibfnamefont
  {S.}~\bibnamefont {Nakamura}}, \bibinfo {author} {\bibfnamefont
  {S.}~\bibnamefont {Awaji}}, \bibinfo {author} {\bibfnamefont
  {A.}~\bibnamefont {Tsukazaki}}, \bibinfo {author} {\bibfnamefont
  {H.}~\bibnamefont {Nakao}}, \emph {et~al.},\ }\bibfield  {title} {\bibinfo
  {title} {Quantum {H}all effect in a bulk antiferromagnet {E}u{M}n{B}i$_2$
  with magnetically confined two-dimensional {D}irac fermions},\ }\href@noop {}
  {\bibfield  {journal} {\bibinfo  {journal} {Sci. Adv.}\ }\textbf {\bibinfo
  {volume} {2}},\ \bibinfo {pages} {e1501117} (\bibinfo {year}
  {2016})}\BibitemShut {NoStop}%
\bibitem [{\citenamefont {Tang}\ \emph {et~al.}(2019)\citenamefont {Tang},
  \citenamefont {Ren}, \citenamefont {Wang}, \citenamefont {Zhong},
  \citenamefont {Schneeloch}, \citenamefont {Yang}, \citenamefont {Yang},
  \citenamefont {Lee}, \citenamefont {Gu}, \citenamefont {Qiao},\ and\
  \citenamefont {Zhang}}]{Tang2019ZrTe5}%
  \BibitemOpen
  \bibfield  {author} {\bibinfo {author} {\bibfnamefont {F.}~\bibnamefont
  {Tang}}, \bibinfo {author} {\bibfnamefont {Y.}~\bibnamefont {Ren}}, \bibinfo
  {author} {\bibfnamefont {P.}~\bibnamefont {Wang}}, \bibinfo {author}
  {\bibfnamefont {R.}~\bibnamefont {Zhong}}, \bibinfo {author} {\bibfnamefont
  {J.}~\bibnamefont {Schneeloch}}, \bibinfo {author} {\bibfnamefont {S.~A.}\
  \bibnamefont {Yang}}, \bibinfo {author} {\bibfnamefont {K.}~\bibnamefont
  {Yang}}, \bibinfo {author} {\bibfnamefont {P.~A.}\ \bibnamefont {Lee}},
  \bibinfo {author} {\bibfnamefont {G.}~\bibnamefont {Gu}}, \bibinfo {author}
  {\bibfnamefont {Z.}~\bibnamefont {Qiao}},\ and\ \bibinfo {author}
  {\bibfnamefont {L.}~\bibnamefont {Zhang}},\ }\bibfield  {title} {\bibinfo
  {title} {Three-dimensional quantum {H}all effect and metal--insulator
  transition in {Z}r{T}e$_5$},\ }\href
  {https://doi.org/10.1038/s41586-019-1180-9} {\bibfield  {journal} {\bibinfo
  {journal} {Nature}\ }\textbf {\bibinfo {volume} {569}},\ \bibinfo {pages}
  {537} (\bibinfo {year} {2019})}\BibitemShut {NoStop}%
\bibitem [{\citenamefont {Goerbig}(2011)}]{Goerbig2011}%
  \BibitemOpen
  \bibfield  {author} {\bibinfo {author} {\bibfnamefont {M.~O.}\ \bibnamefont
  {Goerbig}},\ }\bibfield  {title} {\bibinfo {title} {Electronic properties of
  graphene in a strong magnetic field},\ }\href
  {https://doi.org/10.1103/RevModPhys.83.1193} {\bibfield  {journal} {\bibinfo
  {journal} {Rev. Mod. Phys.}\ }\textbf {\bibinfo {volume} {83}},\ \bibinfo
  {pages} {1193} (\bibinfo {year} {2011})}\BibitemShut {NoStop}%
\bibitem [{\citenamefont {Lv}\ \emph {et~al.}(2021)\citenamefont {Lv},
  \citenamefont {Qian},\ and\ \citenamefont {Ding}}]{Lv2021}%
  \BibitemOpen
  \bibfield  {author} {\bibinfo {author} {\bibfnamefont {B.~Q.}\ \bibnamefont
  {Lv}}, \bibinfo {author} {\bibfnamefont {T.}~\bibnamefont {Qian}},\ and\
  \bibinfo {author} {\bibfnamefont {H.}~\bibnamefont {Ding}},\ }\bibfield
  {title} {\bibinfo {title} {Experimental perspective on three-dimensional
  topological semimetals},\ }\href
  {https://doi.org/10.1103/RevModPhys.93.025002} {\bibfield  {journal}
  {\bibinfo  {journal} {Rev. Mod. Phys.}\ }\textbf {\bibinfo {volume} {93}},\
  \bibinfo {pages} {025002} (\bibinfo {year} {2021})}\BibitemShut {NoStop}%
\bibitem [{\citenamefont {Liu}\ \emph {et~al.}(2016)\citenamefont {Liu},
  \citenamefont {Hu}, \citenamefont {Cao}, \citenamefont {Zhu}, \citenamefont
  {Chuang}, \citenamefont {Graf}, \citenamefont {Adams}, \citenamefont
  {Radmanesh}, \citenamefont {Spinu}, \citenamefont {Chiorescu},\ and\
  \citenamefont {Mao}}]{Liu2016Sb}%
  \BibitemOpen
  \bibfield  {author} {\bibinfo {author} {\bibfnamefont {J.}~\bibnamefont
  {Liu}}, \bibinfo {author} {\bibfnamefont {J.}~\bibnamefont {Hu}}, \bibinfo
  {author} {\bibfnamefont {H.}~\bibnamefont {Cao}}, \bibinfo {author}
  {\bibfnamefont {Y.}~\bibnamefont {Zhu}}, \bibinfo {author} {\bibfnamefont
  {A.}~\bibnamefont {Chuang}}, \bibinfo {author} {\bibfnamefont
  {D.}~\bibnamefont {Graf}}, \bibinfo {author} {\bibfnamefont {D.~J.}\
  \bibnamefont {Adams}}, \bibinfo {author} {\bibfnamefont {S.~M.~A.}\
  \bibnamefont {Radmanesh}}, \bibinfo {author} {\bibfnamefont {L.}~\bibnamefont
  {Spinu}}, \bibinfo {author} {\bibfnamefont {I.}~\bibnamefont {Chiorescu}},\
  and\ \bibinfo {author} {\bibfnamefont {Z.}~\bibnamefont {Mao}},\ }\bibfield
  {title} {\bibinfo {title} {Nearly massless {D}irac fermions hosted by {S}b
  square net in {B}a{M}n{S}b$_2$},\ }\href {https://doi.org/10.1038/srep30525}
  {\bibfield  {journal} {\bibinfo  {journal} {Sci. Rep.}\ }\textbf {\bibinfo
  {volume} {6}},\ \bibinfo {pages} {30525} (\bibinfo {year}
  {2016})}\BibitemShut {NoStop}%
\bibitem [{\citenamefont {Huang}\ \emph {et~al.}(2017)\citenamefont {Huang},
  \citenamefont {Kim}, \citenamefont {Shelton}, \citenamefont {Plummer},\ and\
  \citenamefont {Jin}}]{Huang2017Sb}%
  \BibitemOpen
  \bibfield  {author} {\bibinfo {author} {\bibfnamefont {S.}~\bibnamefont
  {Huang}}, \bibinfo {author} {\bibfnamefont {J.}~\bibnamefont {Kim}}, \bibinfo
  {author} {\bibfnamefont {W.~A.}\ \bibnamefont {Shelton}}, \bibinfo {author}
  {\bibfnamefont {E.~W.}\ \bibnamefont {Plummer}},\ and\ \bibinfo {author}
  {\bibfnamefont {R.}~\bibnamefont {Jin}},\ }\bibfield  {title} {\bibinfo
  {title} {Nontrivial {B}erry phase in magnetic {B}a{M}n{S}b$_2$ semimetal},\
  }\href {https://doi.org/10.1073/pnas.1706657114} {\bibfield  {journal}
  {\bibinfo  {journal} {Proc. Nat. Acad. Sci.}\ }\textbf {\bibinfo {volume}
  {114}},\ \bibinfo {pages} {6256} (\bibinfo {year} {2017})}\BibitemShut
  {NoStop}%
\bibitem [{\citenamefont {Ryu}\ \emph {et~al.}(2018)\citenamefont {Ryu},
  \citenamefont {Park}, \citenamefont {Li}, \citenamefont {Ren}, \citenamefont
  {Neaton}, \citenamefont {Petrovic}, \citenamefont {Hwang},\ and\
  \citenamefont {Mo}}]{ryu2018Zn}%
  \BibitemOpen
  \bibfield  {author} {\bibinfo {author} {\bibfnamefont {H.}~\bibnamefont
  {Ryu}}, \bibinfo {author} {\bibfnamefont {S.~Y.}\ \bibnamefont {Park}},
  \bibinfo {author} {\bibfnamefont {L.}~\bibnamefont {Li}}, \bibinfo {author}
  {\bibfnamefont {W.}~\bibnamefont {Ren}}, \bibinfo {author} {\bibfnamefont
  {J.~B.}\ \bibnamefont {Neaton}}, \bibinfo {author} {\bibfnamefont
  {C.}~\bibnamefont {Petrovic}}, \bibinfo {author} {\bibfnamefont
  {C.}~\bibnamefont {Hwang}},\ and\ \bibinfo {author} {\bibfnamefont {S.-K.}\
  \bibnamefont {Mo}},\ }\bibfield  {title} {\bibinfo {title} {Anisotropic
  {D}irac fermions in {B}a{M}n{B}i$_2$ and {B}a{Z}n{B}i$_2$},\ }\href@noop {}
  {\bibfield  {journal} {\bibinfo  {journal} {Sci. Rep.}\ }\textbf {\bibinfo
  {volume} {8}},\ \bibinfo {pages} {1} (\bibinfo {year} {2018})}\BibitemShut
  {NoStop}%
\bibitem [{\citenamefont {Borisenko}\ \emph {et~al.}(2019)\citenamefont
  {Borisenko}, \citenamefont {Evtushinsky}, \citenamefont {Gibson},
  \citenamefont {Yaresko}, \citenamefont {Koepernik}, \citenamefont {Kim},
  \citenamefont {Ali}, \citenamefont {van~den Brink}, \citenamefont {Hoesch},
  \citenamefont {Fedorov}, \citenamefont {Haubold}, \citenamefont
  {Kushnirenko}, \citenamefont {Soldatov}, \citenamefont {Schafer},\ and\
  \citenamefont {Cava}}]{borisenko2019Yb}%
  \BibitemOpen
  \bibfield  {author} {\bibinfo {author} {\bibfnamefont {S.}~\bibnamefont
  {Borisenko}}, \bibinfo {author} {\bibfnamefont {D.}~\bibnamefont
  {Evtushinsky}}, \bibinfo {author} {\bibfnamefont {Q.}~\bibnamefont {Gibson}},
  \bibinfo {author} {\bibfnamefont {A.}~\bibnamefont {Yaresko}}, \bibinfo
  {author} {\bibfnamefont {K.}~\bibnamefont {Koepernik}}, \bibinfo {author}
  {\bibfnamefont {T.}~\bibnamefont {Kim}}, \bibinfo {author} {\bibfnamefont
  {M.}~\bibnamefont {Ali}}, \bibinfo {author} {\bibfnamefont {J.}~\bibnamefont
  {van~den Brink}}, \bibinfo {author} {\bibfnamefont {M.}~\bibnamefont
  {Hoesch}}, \bibinfo {author} {\bibfnamefont {A.}~\bibnamefont {Fedorov}},
  \bibinfo {author} {\bibfnamefont {E.}~\bibnamefont {Haubold}}, \bibinfo
  {author} {\bibfnamefont {Y.}~\bibnamefont {Kushnirenko}}, \bibinfo {author}
  {\bibfnamefont {I.}~\bibnamefont {Soldatov}}, \bibinfo {author}
  {\bibfnamefont {R.}~\bibnamefont {Schafer}},\ and\ \bibinfo {author}
  {\bibfnamefont {R.~J.}\ \bibnamefont {Cava}},\ }\bibfield  {title} {\bibinfo
  {title} {Time-reversal symmetry breaking type-{II} {W}eyl state in
  {Y}b{M}n{B}i$_2$},\ }\href@noop {} {\bibfield  {journal} {\bibinfo  {journal}
  {Nat. Commun.}\ }\textbf {\bibinfo {volume} {10}},\ \bibinfo {pages} {1}
  (\bibinfo {year} {2019})}\BibitemShut {NoStop}%
\bibitem [{\citenamefont {Klemenz}\ \emph {et~al.}(2019)\citenamefont
  {Klemenz}, \citenamefont {Lei},\ and\ \citenamefont
  {Schoop}}]{Klemenz2019review}%
  \BibitemOpen
  \bibfield  {author} {\bibinfo {author} {\bibfnamefont {S.}~\bibnamefont
  {Klemenz}}, \bibinfo {author} {\bibfnamefont {S.}~\bibnamefont {Lei}},\ and\
  \bibinfo {author} {\bibfnamefont {L.~M.}\ \bibnamefont {Schoop}},\ }\bibfield
   {title} {\bibinfo {title} {Topological semimetals in square-net materials},\
  }\href {https://doi.org/10.1146/annurev-matsci-070218-010114} {\bibfield
  {journal} {\bibinfo  {journal} {Ann. Rev. Mater. Research}\ }\textbf
  {\bibinfo {volume} {49}},\ \bibinfo {pages} {185} (\bibinfo {year}
  {2019})}\BibitemShut {NoStop}%
\bibitem [{\citenamefont {Li}\ \emph {et~al.}(2016)\citenamefont {Li},
  \citenamefont {Wang}, \citenamefont {Graf}, \citenamefont {Wang},
  \citenamefont {Wang}, \citenamefont {Petrovic} \emph {et~al.}}]{li2016Bi}%
  \BibitemOpen
  \bibfield  {author} {\bibinfo {author} {\bibfnamefont {L.}~\bibnamefont
  {Li}}, \bibinfo {author} {\bibfnamefont {K.}~\bibnamefont {Wang}}, \bibinfo
  {author} {\bibfnamefont {D.}~\bibnamefont {Graf}}, \bibinfo {author}
  {\bibfnamefont {L.}~\bibnamefont {Wang}}, \bibinfo {author} {\bibfnamefont
  {A.}~\bibnamefont {Wang}}, \bibinfo {author} {\bibfnamefont {C.}~\bibnamefont
  {Petrovic}}, \emph {et~al.},\ }\bibfield  {title} {\bibinfo {title}
  {Electron-hole asymmetry, {D}irac fermions, and quantum magnetoresistance in
  {B}a{M}n{B}i$_2$},\ }\href {https://doi.org/10.1103/PhysRevB.93.115141}
  {\bibfield  {journal} {\bibinfo  {journal} {Phys. Rev. B}\ }\textbf {\bibinfo
  {volume} {93}},\ \bibinfo {pages} {115141} (\bibinfo {year}
  {2016})}\BibitemShut {NoStop}%
\bibitem [{\citenamefont {Sakai}\ \emph {et~al.}(2020)\citenamefont {Sakai},
  \citenamefont {Fujimura}, \citenamefont {Sakuragi}, \citenamefont {Ochi},
  \citenamefont {Kurihara}, \citenamefont {Miyake}, \citenamefont {Tokunaga},
  \citenamefont {Kojima}, \citenamefont {Hashizume}, \citenamefont {Muro},
  \citenamefont {Kuroda}, \citenamefont {Kondo}, \citenamefont {Kida},
  \citenamefont {Hagiwara}, \citenamefont {Kuroki}, \citenamefont {Kondo},
  \citenamefont {Tsuruda}, \citenamefont {Murakawa},\ and\ \citenamefont
  {Hanasaki}}]{sakai2020Sb}%
  \BibitemOpen
  \bibfield  {author} {\bibinfo {author} {\bibfnamefont {H.}~\bibnamefont
  {Sakai}}, \bibinfo {author} {\bibfnamefont {H.}~\bibnamefont {Fujimura}},
  \bibinfo {author} {\bibfnamefont {S.}~\bibnamefont {Sakuragi}}, \bibinfo
  {author} {\bibfnamefont {M.}~\bibnamefont {Ochi}}, \bibinfo {author}
  {\bibfnamefont {R.}~\bibnamefont {Kurihara}}, \bibinfo {author}
  {\bibfnamefont {A.}~\bibnamefont {Miyake}}, \bibinfo {author} {\bibfnamefont
  {M.}~\bibnamefont {Tokunaga}}, \bibinfo {author} {\bibfnamefont
  {T.}~\bibnamefont {Kojima}}, \bibinfo {author} {\bibfnamefont
  {D.}~\bibnamefont {Hashizume}}, \bibinfo {author} {\bibfnamefont
  {T.}~\bibnamefont {Muro}}, \bibinfo {author} {\bibfnamefont {K.}~\bibnamefont
  {Kuroda}}, \bibinfo {author} {\bibfnamefont {T.}~\bibnamefont {Kondo}},
  \bibinfo {author} {\bibfnamefont {T.}~\bibnamefont {Kida}}, \bibinfo {author}
  {\bibfnamefont {M.}~\bibnamefont {Hagiwara}}, \bibinfo {author}
  {\bibfnamefont {K.}~\bibnamefont {Kuroki}}, \bibinfo {author} {\bibfnamefont
  {M.}~\bibnamefont {Kondo}}, \bibinfo {author} {\bibfnamefont
  {K.}~\bibnamefont {Tsuruda}}, \bibinfo {author} {\bibfnamefont
  {H.}~\bibnamefont {Murakawa}},\ and\ \bibinfo {author} {\bibfnamefont
  {N.}~\bibnamefont {Hanasaki}},\ }\bibfield  {title} {\bibinfo {title} {Bulk
  quantum {H}all effect of spin-valley coupled {D}irac fermions in the polar
  antiferromagnet {B}a{M}n{Sb}$_{2}$},\ }\href
  {https://doi.org/10.1103/PhysRevB.101.081104} {\bibfield  {journal} {\bibinfo
   {journal} {Phys. Rev. B}\ }\textbf {\bibinfo {volume} {101}},\ \bibinfo
  {pages} {081104} (\bibinfo {year} {2020})}\BibitemShut {NoStop}%
\bibitem [{\citenamefont {Liu}\ \emph {et~al.}(2021)\citenamefont {Liu},
  \citenamefont {Yu}, \citenamefont {Ning}, \citenamefont {Yi}, \citenamefont
  {Miao}, \citenamefont {Min}, \citenamefont {Zhao}, \citenamefont {Ning},
  \citenamefont {Lopez}, \citenamefont {Zhu}, \citenamefont {Pillsbury},
  \citenamefont {Zhang}, \citenamefont {Wang}, \citenamefont {Hu},
  \citenamefont {Cao}, \citenamefont {Chakoumakos}, \citenamefont {Balakirev},
  \citenamefont {Weickert}, \citenamefont {Jaime}, \citenamefont {Lai},
  \citenamefont {Yang}, \citenamefont {Sun}, \citenamefont {Alem},
  \citenamefont {Gopalan}, \citenamefont {Chang}, \citenamefont {Samarth},
  \citenamefont {Liu}, \citenamefont {McDonald},\ and\ \citenamefont
  {Mao}}]{Liu2021Sb}%
  \BibitemOpen
  \bibfield  {author} {\bibinfo {author} {\bibfnamefont {J.}~\bibnamefont
  {Liu}}, \bibinfo {author} {\bibfnamefont {J.}~\bibnamefont {Yu}}, \bibinfo
  {author} {\bibfnamefont {J.~L.}\ \bibnamefont {Ning}}, \bibinfo {author}
  {\bibfnamefont {H.~M.}\ \bibnamefont {Yi}}, \bibinfo {author} {\bibfnamefont
  {L.}~\bibnamefont {Miao}}, \bibinfo {author} {\bibfnamefont {L.~J.}\
  \bibnamefont {Min}}, \bibinfo {author} {\bibfnamefont {Y.~F.}\ \bibnamefont
  {Zhao}}, \bibinfo {author} {\bibfnamefont {W.}~\bibnamefont {Ning}}, \bibinfo
  {author} {\bibfnamefont {K.~A.}\ \bibnamefont {Lopez}}, \bibinfo {author}
  {\bibfnamefont {Y.~L.}\ \bibnamefont {Zhu}}, \bibinfo {author} {\bibfnamefont
  {T.}~\bibnamefont {Pillsbury}}, \bibinfo {author} {\bibfnamefont {Y.~B.}\
  \bibnamefont {Zhang}}, \bibinfo {author} {\bibfnamefont {Y.}~\bibnamefont
  {Wang}}, \bibinfo {author} {\bibfnamefont {J.}~\bibnamefont {Hu}}, \bibinfo
  {author} {\bibfnamefont {H.~B.}\ \bibnamefont {Cao}}, \bibinfo {author}
  {\bibfnamefont {B.~C.}\ \bibnamefont {Chakoumakos}}, \bibinfo {author}
  {\bibfnamefont {F.}~\bibnamefont {Balakirev}}, \bibinfo {author}
  {\bibfnamefont {F.}~\bibnamefont {Weickert}}, \bibinfo {author}
  {\bibfnamefont {M.}~\bibnamefont {Jaime}}, \bibinfo {author} {\bibfnamefont
  {Y.}~\bibnamefont {Lai}}, \bibinfo {author} {\bibfnamefont {K.}~\bibnamefont
  {Yang}}, \bibinfo {author} {\bibfnamefont {J.~W.}\ \bibnamefont {Sun}},
  \bibinfo {author} {\bibfnamefont {N.}~\bibnamefont {Alem}}, \bibinfo {author}
  {\bibfnamefont {V.}~\bibnamefont {Gopalan}}, \bibinfo {author} {\bibfnamefont
  {C.~Z.}\ \bibnamefont {Chang}}, \bibinfo {author} {\bibfnamefont
  {N.}~\bibnamefont {Samarth}}, \bibinfo {author} {\bibfnamefont {C.~X.}\
  \bibnamefont {Liu}}, \bibinfo {author} {\bibfnamefont {R.~D.}\ \bibnamefont
  {McDonald}},\ and\ \bibinfo {author} {\bibfnamefont {Z.~Q.}\ \bibnamefont
  {Mao}},\ }\bibfield  {title} {\bibinfo {title} {Spin-valley locking and bulk
  quantum {H}all effect in a noncentrosymmetric {D}irac semimetal
  {B}a{M}n{S}b$_2$},\ }\href {https://doi.org/10.1038/s41467-021-24369-1}
  {\bibfield  {journal} {\bibinfo  {journal} {Nat. Commun.}\ }\textbf {\bibinfo
  {volume} {12}},\ \bibinfo {pages} {4062} (\bibinfo {year}
  {2021})}\BibitemShut {NoStop}%
\bibitem [{\citenamefont {Kondo}\ \emph {et~al.}(2021)\citenamefont {Kondo},
  \citenamefont {Ochi}, \citenamefont {Kojima}, \citenamefont {Kurihara},
  \citenamefont {Sekine}, \citenamefont {Matsubara}, \citenamefont {Miyake},
  \citenamefont {Tokunaga}, \citenamefont {Kuroki}, \citenamefont {Murakawa},
  \citenamefont {Hanasaki},\ and\ \citenamefont {Sakai}}]{Kondo2021Bi}%
  \BibitemOpen
  \bibfield  {author} {\bibinfo {author} {\bibfnamefont {M.}~\bibnamefont
  {Kondo}}, \bibinfo {author} {\bibfnamefont {M.}~\bibnamefont {Ochi}},
  \bibinfo {author} {\bibfnamefont {T.}~\bibnamefont {Kojima}}, \bibinfo
  {author} {\bibfnamefont {R.}~\bibnamefont {Kurihara}}, \bibinfo {author}
  {\bibfnamefont {D.}~\bibnamefont {Sekine}}, \bibinfo {author} {\bibfnamefont
  {M.}~\bibnamefont {Matsubara}}, \bibinfo {author} {\bibfnamefont
  {A.}~\bibnamefont {Miyake}}, \bibinfo {author} {\bibfnamefont
  {M.}~\bibnamefont {Tokunaga}}, \bibinfo {author} {\bibfnamefont
  {K.}~\bibnamefont {Kuroki}}, \bibinfo {author} {\bibfnamefont
  {H.}~\bibnamefont {Murakawa}}, \bibinfo {author} {\bibfnamefont
  {N.}~\bibnamefont {Hanasaki}},\ and\ \bibinfo {author} {\bibfnamefont
  {H.}~\bibnamefont {Sakai}},\ }\bibfield  {title} {\bibinfo {title} {Tunable
  spin-valley coupling in layered polar {D}irac metals},\ }\href
  {https://doi.org/10.1038/s43246-021-00152-z} {\bibfield  {journal} {\bibinfo
  {journal} {Comm. Mater.}\ }\textbf {\bibinfo {volume} {2}},\ \bibinfo {pages}
  {49} (\bibinfo {year} {2021})}\BibitemShut {NoStop}%
\bibitem [{\citenamefont {Ma}\ \emph {et~al.}(2019)\citenamefont {Ma},
  \citenamefont {Xu}, \citenamefont {Shen}, \citenamefont {MacNeill},
  \citenamefont {Fatemi}, \citenamefont {Chang}, \citenamefont {Mier~Valdivia},
  \citenamefont {Wu}, \citenamefont {Du}, \citenamefont {Hsu}, \citenamefont
  {Fang}, \citenamefont {Gibson}, \citenamefont {Watanabe}, \citenamefont
  {Taniguchi}, \citenamefont {Cava}, \citenamefont {Kaxiras}, \citenamefont
  {Lu}, \citenamefont {Lin}, \citenamefont {Fu}, \citenamefont {Gedik},\ and\
  \citenamefont {Jarillo-Herrero}}]{Ma2019}%
  \BibitemOpen
  \bibfield  {author} {\bibinfo {author} {\bibfnamefont {Q.}~\bibnamefont
  {Ma}}, \bibinfo {author} {\bibfnamefont {S.-Y.}\ \bibnamefont {Xu}}, \bibinfo
  {author} {\bibfnamefont {H.}~\bibnamefont {Shen}}, \bibinfo {author}
  {\bibfnamefont {D.}~\bibnamefont {MacNeill}}, \bibinfo {author}
  {\bibfnamefont {V.}~\bibnamefont {Fatemi}}, \bibinfo {author} {\bibfnamefont
  {T.-R.}\ \bibnamefont {Chang}}, \bibinfo {author} {\bibfnamefont {A.~M.}\
  \bibnamefont {Mier~Valdivia}}, \bibinfo {author} {\bibfnamefont
  {S.}~\bibnamefont {Wu}}, \bibinfo {author} {\bibfnamefont {Z.}~\bibnamefont
  {Du}}, \bibinfo {author} {\bibfnamefont {C.-H.}\ \bibnamefont {Hsu}},
  \bibinfo {author} {\bibfnamefont {S.}~\bibnamefont {Fang}}, \bibinfo {author}
  {\bibfnamefont {Q.~D.}\ \bibnamefont {Gibson}}, \bibinfo {author}
  {\bibfnamefont {K.}~\bibnamefont {Watanabe}}, \bibinfo {author}
  {\bibfnamefont {T.}~\bibnamefont {Taniguchi}}, \bibinfo {author}
  {\bibfnamefont {R.~J.}\ \bibnamefont {Cava}}, \bibinfo {author}
  {\bibfnamefont {E.}~\bibnamefont {Kaxiras}}, \bibinfo {author} {\bibfnamefont
  {H.-Z.}\ \bibnamefont {Lu}}, \bibinfo {author} {\bibfnamefont
  {H.}~\bibnamefont {Lin}}, \bibinfo {author} {\bibfnamefont {L.}~\bibnamefont
  {Fu}}, \bibinfo {author} {\bibfnamefont {N.}~\bibnamefont {Gedik}},\ and\
  \bibinfo {author} {\bibfnamefont {P.}~\bibnamefont {Jarillo-Herrero}},\
  }\bibfield  {title} {\bibinfo {title} {Observation of the nonlinear {H}all
  effect under time-reversal-symmetric conditions},\ }\href
  {https://doi.org/10.1038/s41586-018-0807-6} {\bibfield  {journal} {\bibinfo
  {journal} {Nature}\ }\textbf {\bibinfo {volume} {565}},\ \bibinfo {pages}
  {337} (\bibinfo {year} {2019})}\BibitemShut {NoStop}%
\bibitem [{\citenamefont {Lee}\ \emph {et~al.}(2013)\citenamefont {Lee},
  \citenamefont {Farhan}, \citenamefont {Kim},\ and\ \citenamefont
  {Shim}}]{lee2013Sr}%
  \BibitemOpen
  \bibfield  {author} {\bibinfo {author} {\bibfnamefont {G.}~\bibnamefont
  {Lee}}, \bibinfo {author} {\bibfnamefont {M.~A.}\ \bibnamefont {Farhan}},
  \bibinfo {author} {\bibfnamefont {J.~S.}\ \bibnamefont {Kim}},\ and\ \bibinfo
  {author} {\bibfnamefont {J.~H.}\ \bibnamefont {Shim}},\ }\bibfield  {title}
  {\bibinfo {title} {Anisotropic {D}irac electronic structures of
  ${A}${M}n{B}i$_{2}$ (${A}=\mathrm{Sr}$,{C}a)},\ }\href
  {https://doi.org/10.1103/PhysRevB.87.245104} {\bibfield  {journal} {\bibinfo
  {journal} {Phys. Rev. B}\ }\textbf {\bibinfo {volume} {87}},\ \bibinfo
  {pages} {245104} (\bibinfo {year} {2013})}\BibitemShut {NoStop}%
\bibitem [{\citenamefont {Hirata}\ \emph {et~al.}(2017)\citenamefont {Hirata},
  \citenamefont {Ishikawa}, \citenamefont {Matsuno}, \citenamefont {Kobayashi},
  \citenamefont {Miyagawa}, \citenamefont {Tamura}, \citenamefont {Berthier},\
  and\ \citenamefont {Kanoda}}]{hirata2017I3}%
  \BibitemOpen
  \bibfield  {author} {\bibinfo {author} {\bibfnamefont {M.}~\bibnamefont
  {Hirata}}, \bibinfo {author} {\bibfnamefont {K.}~\bibnamefont {Ishikawa}},
  \bibinfo {author} {\bibfnamefont {G.}~\bibnamefont {Matsuno}}, \bibinfo
  {author} {\bibfnamefont {A.}~\bibnamefont {Kobayashi}}, \bibinfo {author}
  {\bibfnamefont {K.}~\bibnamefont {Miyagawa}}, \bibinfo {author}
  {\bibfnamefont {M.}~\bibnamefont {Tamura}}, \bibinfo {author} {\bibfnamefont
  {C.}~\bibnamefont {Berthier}},\ and\ \bibinfo {author} {\bibfnamefont
  {K.}~\bibnamefont {Kanoda}},\ }\bibfield  {title} {\bibinfo {title}
  {Anomalous spin correlations and excitonic instability of interacting 2{D}
  {W}eyl fermions},\ }\href@noop {} {\bibfield  {journal} {\bibinfo  {journal}
  {Science}\ }\textbf {\bibinfo {volume} {358}},\ \bibinfo {pages} {1403}
  (\bibinfo {year} {2017})}\BibitemShut {NoStop}%
\bibitem [{\citenamefont {Nisson}\ \emph {et~al.}(2013)\citenamefont {Nisson},
  \citenamefont {Dioguardi}, \citenamefont {Klavins}, \citenamefont {Lin},
  \citenamefont {Shirer}, \citenamefont {Shockley}, \citenamefont {Crocker},\
  and\ \citenamefont {Curro}}]{nisson2013nuclear}%
  \BibitemOpen
  \bibfield  {author} {\bibinfo {author} {\bibfnamefont {D.}~\bibnamefont
  {Nisson}}, \bibinfo {author} {\bibfnamefont {A.}~\bibnamefont {Dioguardi}},
  \bibinfo {author} {\bibfnamefont {P.}~\bibnamefont {Klavins}}, \bibinfo
  {author} {\bibfnamefont {C.}~\bibnamefont {Lin}}, \bibinfo {author}
  {\bibfnamefont {K.}~\bibnamefont {Shirer}}, \bibinfo {author} {\bibfnamefont
  {A.}~\bibnamefont {Shockley}}, \bibinfo {author} {\bibfnamefont
  {J.}~\bibnamefont {Crocker}},\ and\ \bibinfo {author} {\bibfnamefont
  {N.}~\bibnamefont {Curro}},\ }\bibfield  {title} {\bibinfo {title} {Nuclear
  magnetic resonance as a probe of electronic states of {B}i$_2${S}e$_3$},\
  }\href@noop {} {\bibfield  {journal} {\bibinfo  {journal} {Phys. Rev. B}\
  }\textbf {\bibinfo {volume} {87}},\ \bibinfo {pages} {195202} (\bibinfo
  {year} {2013})}\BibitemShut {NoStop}%
\bibitem [{\citenamefont {Yasuoka}\ \emph {et~al.}(2017)\citenamefont
  {Yasuoka}, \citenamefont {Kubo}, \citenamefont {Kishimoto}, \citenamefont
  {Kasinathan}, \citenamefont {Schmidt}, \citenamefont {Yan}, \citenamefont
  {Zhang}, \citenamefont {Tou}, \citenamefont {Felser}, \citenamefont
  {Mackenzie} \emph {et~al.}}]{yasuoka2017Weyl}%
  \BibitemOpen
  \bibfield  {author} {\bibinfo {author} {\bibfnamefont {H.}~\bibnamefont
  {Yasuoka}}, \bibinfo {author} {\bibfnamefont {T.}~\bibnamefont {Kubo}},
  \bibinfo {author} {\bibfnamefont {Y.}~\bibnamefont {Kishimoto}}, \bibinfo
  {author} {\bibfnamefont {D.}~\bibnamefont {Kasinathan}}, \bibinfo {author}
  {\bibfnamefont {M.}~\bibnamefont {Schmidt}}, \bibinfo {author} {\bibfnamefont
  {B.}~\bibnamefont {Yan}}, \bibinfo {author} {\bibfnamefont {Y.}~\bibnamefont
  {Zhang}}, \bibinfo {author} {\bibfnamefont {H.}~\bibnamefont {Tou}}, \bibinfo
  {author} {\bibfnamefont {C.}~\bibnamefont {Felser}}, \bibinfo {author}
  {\bibfnamefont {A.}~\bibnamefont {Mackenzie}}, \emph {et~al.},\ }\bibfield
  {title} {\bibinfo {title} {Emergent {W}eyl fermion excitations in {T}a{P}
  explored by $^{181}${T}a quadrupole resonance},\ }\href@noop {} {\bibfield
  {journal} {\bibinfo  {journal} {Phys. Rev. Lett.}\ }\textbf {\bibinfo
  {volume} {118}},\ \bibinfo {pages} {236403} (\bibinfo {year}
  {2017})}\BibitemShut {NoStop}%
\bibitem [{\citenamefont {Tian}\ \emph {et~al.}(2019)\citenamefont {Tian},
  \citenamefont {Ghassemi},\ and\ \citenamefont {Ross}}]{Tian2019}%
  \BibitemOpen
  \bibfield  {author} {\bibinfo {author} {\bibfnamefont {Y.}~\bibnamefont
  {Tian}}, \bibinfo {author} {\bibfnamefont {N.}~\bibnamefont {Ghassemi}},\
  and\ \bibinfo {author} {\bibfnamefont {J.~H.}\ \bibnamefont {Ross}},\
  }\bibfield  {title} {\bibinfo {title} {Dirac electron behavior and {NMR}
  evidence for topological band inversion in {Z}r{T}e$_{5}$},\ }\href
  {https://doi.org/10.1103/PhysRevB.100.165149} {\bibfield  {journal} {\bibinfo
   {journal} {Phys. Rev. B}\ }\textbf {\bibinfo {volume} {100}},\ \bibinfo
  {pages} {165149} (\bibinfo {year} {2019})}\BibitemShut {NoStop}%
\bibitem [{\citenamefont {Wang}\ \emph {et~al.}(2020)\citenamefont {Wang},
  \citenamefont {Honjo}, \citenamefont {Zhao}, \citenamefont {Chen},
  \citenamefont {Matano}, \citenamefont {Zhou},\ and\ \citenamefont
  {Zheng}}]{wang2020As}%
  \BibitemOpen
  \bibfield  {author} {\bibinfo {author} {\bibfnamefont {C.}~\bibnamefont
  {Wang}}, \bibinfo {author} {\bibfnamefont {Y.}~\bibnamefont {Honjo}},
  \bibinfo {author} {\bibfnamefont {L.}~\bibnamefont {Zhao}}, \bibinfo {author}
  {\bibfnamefont {G.}~\bibnamefont {Chen}}, \bibinfo {author} {\bibfnamefont
  {K.}~\bibnamefont {Matano}}, \bibinfo {author} {\bibfnamefont
  {R.}~\bibnamefont {Zhou}},\ and\ \bibinfo {author} {\bibfnamefont {G.-q.}\
  \bibnamefont {Zheng}},\ }\bibfield  {title} {\bibinfo {title} {Landau
  diamagnetism and {W}eyl-fermion excitations in {T}a{A}s revealed by
  $^{75}${A}s {NMR} and {NQR}},\ }\href@noop {} {\bibfield  {journal} {\bibinfo
   {journal} {Phys. Rev. B}\ }\textbf {\bibinfo {volume} {101}},\ \bibinfo
  {pages} {241110} (\bibinfo {year} {2020})}\BibitemShut {NoStop}%
\bibitem [{\citenamefont {Tian}\ \emph {et~al.}(2021)\citenamefont {Tian},
  \citenamefont {Ghassemi},\ and\ \citenamefont {Ross}}]{Tian2021ZrTe5}%
  \BibitemOpen
  \bibfield  {author} {\bibinfo {author} {\bibfnamefont {Y.}~\bibnamefont
  {Tian}}, \bibinfo {author} {\bibfnamefont {N.}~\bibnamefont {Ghassemi}},\
  and\ \bibinfo {author} {\bibfnamefont {J.~H.}\ \bibnamefont {Ross}},\
  }\bibfield  {title} {\bibinfo {title} {Gap-opening transition in {D}irac
  semimetal {Z}r{T}e$_{5}$},\ }\href
  {https://doi.org/10.1103/PhysRevLett.126.236401} {\bibfield  {journal}
  {\bibinfo  {journal} {Phys. Rev. Lett.}\ }\textbf {\bibinfo {volume} {126}},\
  \bibinfo {pages} {236401} (\bibinfo {year} {2021})}\BibitemShut {NoStop}%
\bibitem [{\citenamefont {Papawassiliou}\ \emph {et~al.}(2020)\citenamefont
  {Papawassiliou}, \citenamefont {Jaworski}, \citenamefont {Pell},
  \citenamefont {Jang}, \citenamefont {Kim}, \citenamefont {Lee}, \citenamefont
  {Kim}, \citenamefont {Alwahedi}, \citenamefont {Alhassan}, \citenamefont
  {Subrati}, \citenamefont {Fardis}, \citenamefont {Karagianni}, \citenamefont
  {Panopoulos}, \citenamefont {Dolinsek},\ and\ \citenamefont
  {Papavassiliou}}]{Papawassiliou2020}%
  \BibitemOpen
  \bibfield  {author} {\bibinfo {author} {\bibfnamefont {W.}~\bibnamefont
  {Papawassiliou}}, \bibinfo {author} {\bibfnamefont {A.}~\bibnamefont
  {Jaworski}}, \bibinfo {author} {\bibfnamefont {A.~J.}\ \bibnamefont {Pell}},
  \bibinfo {author} {\bibfnamefont {J.~H.}\ \bibnamefont {Jang}}, \bibinfo
  {author} {\bibfnamefont {Y.}~\bibnamefont {Kim}}, \bibinfo {author}
  {\bibfnamefont {S.-C.}\ \bibnamefont {Lee}}, \bibinfo {author} {\bibfnamefont
  {H.~J.}\ \bibnamefont {Kim}}, \bibinfo {author} {\bibfnamefont
  {Y.}~\bibnamefont {Alwahedi}}, \bibinfo {author} {\bibfnamefont
  {S.}~\bibnamefont {Alhassan}}, \bibinfo {author} {\bibfnamefont
  {A.}~\bibnamefont {Subrati}}, \bibinfo {author} {\bibfnamefont
  {M.}~\bibnamefont {Fardis}}, \bibinfo {author} {\bibfnamefont
  {M.}~\bibnamefont {Karagianni}}, \bibinfo {author} {\bibfnamefont
  {N.}~\bibnamefont {Panopoulos}}, \bibinfo {author} {\bibfnamefont
  {J.}~\bibnamefont {Dolinsek}},\ and\ \bibinfo {author} {\bibfnamefont
  {G.}~\bibnamefont {Papavassiliou}},\ }\bibfield  {title} {\bibinfo {title}
  {Resolving {D}irac electrons with broadband high-resolution {NMR}},\ }\href
  {https://doi.org/10.1038/s41467-020-14838-4} {\bibfield  {journal} {\bibinfo
  {journal} {Nat. Commun.}\ }\textbf {\bibinfo {volume} {11}},\ \bibinfo
  {pages} {1285} (\bibinfo {year} {2020})}\BibitemShut {NoStop}%
\bibitem [{\citenamefont {Watanabe}\ \emph {et~al.}(2021)\citenamefont
  {Watanabe}, \citenamefont {Kumazaki}, \citenamefont {Ezure}, \citenamefont
  {Sasagawa}, \citenamefont {Cava}, \citenamefont {Itoh},\ and\ \citenamefont
  {Shimizu}}]{Watanabe}%
  \BibitemOpen
  \bibfield  {author} {\bibinfo {author} {\bibfnamefont {Y.}~\bibnamefont
  {Watanabe}}, \bibinfo {author} {\bibfnamefont {M.}~\bibnamefont {Kumazaki}},
  \bibinfo {author} {\bibfnamefont {H.}~\bibnamefont {Ezure}}, \bibinfo
  {author} {\bibfnamefont {T.}~\bibnamefont {Sasagawa}}, \bibinfo {author}
  {\bibfnamefont {R.}~\bibnamefont {Cava}}, \bibinfo {author} {\bibfnamefont
  {M.}~\bibnamefont {Itoh}},\ and\ \bibinfo {author} {\bibfnamefont
  {Y.}~\bibnamefont {Shimizu}},\ }\bibfield  {title} {\bibinfo {title} {Local
  observations of orbital diamagnetism and excitation in three-dimensional
  {D}irac fermion systems {B}i$_{1-x}${S}b$_x$},\ }\href
  {https://doi.org/10.7566/JPSJ.90.053701} {\bibfield  {journal} {\bibinfo
  {journal} {J. Phys. Soc. Jpn.}\ }\textbf {\bibinfo {volume} {90}},\ \bibinfo
  {pages} {053701} (\bibinfo {year} {2021})}\BibitemShut {NoStop}%
\bibitem [{\citenamefont {Hirata}\ \emph {et~al.}(2021)\citenamefont {Hirata},
  \citenamefont {Kobayashi}, \citenamefont {Berthier},\ and\ \citenamefont
  {Kanoda}}]{Hirata2021}%
  \BibitemOpen
  \bibfield  {author} {\bibinfo {author} {\bibfnamefont {M.}~\bibnamefont
  {Hirata}}, \bibinfo {author} {\bibfnamefont {A.}~\bibnamefont {Kobayashi}},
  \bibinfo {author} {\bibfnamefont {C.}~\bibnamefont {Berthier}},\ and\
  \bibinfo {author} {\bibfnamefont {K.}~\bibnamefont {Kanoda}},\ }\bibfield
  {title} {\bibinfo {title} {Interacting chiral electrons at the 2{D} {D}irac
  points: a review},\ }\href {https://doi.org/10.1088/1361-6633/abc17c}
  {\bibfield  {journal} {\bibinfo  {journal} {Rep. Prog. Phys.}\ }\textbf
  {\bibinfo {volume} {84}},\ \bibinfo {pages} {036502} (\bibinfo {year}
  {2021})}\BibitemShut {NoStop}%
\bibitem [{\citenamefont {Yokoo}\ \emph {et~al.}(2022)\citenamefont {Yokoo},
  \citenamefont {Watanabe}, \citenamefont {Kumazaki}, \citenamefont {Itoh},\
  and\ \citenamefont {Shimizu}}]{Yokoo2022}%
  \BibitemOpen
  \bibfield  {author} {\bibinfo {author} {\bibfnamefont {T.}~\bibnamefont
  {Yokoo}}, \bibinfo {author} {\bibfnamefont {Y.}~\bibnamefont {Watanabe}},
  \bibinfo {author} {\bibfnamefont {M.}~\bibnamefont {Kumazaki}}, \bibinfo
  {author} {\bibfnamefont {M.}~\bibnamefont {Itoh}},\ and\ \bibinfo {author}
  {\bibfnamefont {Y.}~\bibnamefont {Shimizu}},\ }\bibfield  {title} {\bibinfo
  {title} {Site-dependent local spin susceptibility and low-energy excitation
  in a weyl semimetal {WT}e$_2$},\ }\href
  {https://doi.org/10.7566/JPSJ.91.054701} {\bibfield  {journal} {\bibinfo
  {journal} {J. Phys. Soc. Jpn.}\ }\textbf {\bibinfo {volume} {91}},\ \bibinfo
  {pages} {054701} (\bibinfo {year} {2022})}\BibitemShut {NoStop}%
\bibitem [{\citenamefont {Hirosawa}\ \emph {et~al.}(2017)\citenamefont
  {Hirosawa}, \citenamefont {Maebashi},\ and\ \citenamefont
  {Ogata}}]{hirosawa2017T1}%
  \BibitemOpen
  \bibfield  {author} {\bibinfo {author} {\bibfnamefont {T.}~\bibnamefont
  {Hirosawa}}, \bibinfo {author} {\bibfnamefont {H.}~\bibnamefont {Maebashi}},\
  and\ \bibinfo {author} {\bibfnamefont {M.}~\bibnamefont {Ogata}},\ }\bibfield
   {title} {\bibinfo {title} {Nuclear spin relaxation time due to the orbital
  currents in {D}irac electron systems},\ }\href
  {https://doi.org/10.7566/JPSJ.86.063705} {\bibfield  {journal} {\bibinfo
  {journal} {J. Phys. Soc. Jpn.}\ }\textbf {\bibinfo {volume} {86}},\ \bibinfo
  {pages} {063705} (\bibinfo {year} {2017})}\BibitemShut {NoStop}%
\bibitem [{\citenamefont {D{\'o}ra}\ and\ \citenamefont
  {Simon}(2009)}]{dora2009graphene}%
  \BibitemOpen
  \bibfield  {author} {\bibinfo {author} {\bibfnamefont {B.}~\bibnamefont
  {D{\'o}ra}}\ and\ \bibinfo {author} {\bibfnamefont {F.}~\bibnamefont
  {Simon}},\ }\bibfield  {title} {\bibinfo {title} {Unusual hyperfine
  interaction of {D}irac electrons and {NMR} spectroscopy in graphene},\
  }\href@noop {} {\bibfield  {journal} {\bibinfo  {journal} {Phys. Rev. Lett.}\
  }\textbf {\bibinfo {volume} {102}},\ \bibinfo {pages} {197602} (\bibinfo
  {year} {2009})}\BibitemShut {NoStop}%
\bibitem [{\citenamefont {Katayama}\ \emph {et~al.}(2006)\citenamefont
  {Katayama}, \citenamefont {Kobayashi},\ and\ \citenamefont
  {Suzumura}}]{Katayama}%
  \BibitemOpen
  \bibfield  {author} {\bibinfo {author} {\bibfnamefont {S.}~\bibnamefont
  {Katayama}}, \bibinfo {author} {\bibfnamefont {A.}~\bibnamefont
  {Kobayashi}},\ and\ \bibinfo {author} {\bibfnamefont {Y.}~\bibnamefont
  {Suzumura}},\ }\bibfield  {title} {\bibinfo {title} {Pressure-induced
  zero-gap semiconducting state in organic conductor
  $\alpha$-({BEDT-TTF})$_2${I}$_3$ salt},\ }\href
  {https://doi.org/10.1143/JPSJ.75.054705} {\bibfield  {journal} {\bibinfo
  {journal} {J. Phys. Soc. Jpn.}\ }\textbf {\bibinfo {volume} {75}},\ \bibinfo
  {pages} {054705} (\bibinfo {year} {2006})}\BibitemShut {NoStop}%
\bibitem [{\citenamefont {Okv{\'a}tovity}\ \emph {et~al.}(2019)\citenamefont
  {Okv{\'a}tovity}, \citenamefont {Yasuoka}, \citenamefont {Baenitz},
  \citenamefont {Simon},\ and\ \citenamefont {D{\'o}ra}}]{okvatovity2019TaP}%
  \BibitemOpen
  \bibfield  {author} {\bibinfo {author} {\bibfnamefont {Z.}~\bibnamefont
  {Okv{\'a}tovity}}, \bibinfo {author} {\bibfnamefont {H.}~\bibnamefont
  {Yasuoka}}, \bibinfo {author} {\bibfnamefont {M.}~\bibnamefont {Baenitz}},
  \bibinfo {author} {\bibfnamefont {F.}~\bibnamefont {Simon}},\ and\ \bibinfo
  {author} {\bibfnamefont {B.}~\bibnamefont {D{\'o}ra}},\ }\bibfield  {title}
  {\bibinfo {title} {Nuclear spin-lattice relaxation time in {T}a{P} and the
  {K}night shift of {W}eyl semimetals},\ }\href@noop {} {\bibfield  {journal}
  {\bibinfo  {journal} {Phys. Rev. B}\ }\textbf {\bibinfo {volume} {99}},\
  \bibinfo {pages} {115107} (\bibinfo {year} {2019})}\BibitemShut {NoStop}%
\bibitem [{\citenamefont {Maebashi}\ \emph {et~al.}(2019)\citenamefont
  {Maebashi}, \citenamefont {Hirosawa}, \citenamefont {Ogata},\ and\
  \citenamefont {Fukuyama}}]{Maebashi2019}%
  \BibitemOpen
  \bibfield  {author} {\bibinfo {author} {\bibfnamefont {H.}~\bibnamefont
  {Maebashi}}, \bibinfo {author} {\bibfnamefont {T.}~\bibnamefont {Hirosawa}},
  \bibinfo {author} {\bibfnamefont {M.}~\bibnamefont {Ogata}},\ and\ \bibinfo
  {author} {\bibfnamefont {H.}~\bibnamefont {Fukuyama}},\ }\bibfield  {title}
  {\bibinfo {title} {Nuclear magnetic relaxation and knight shift due to
  orbital interaction in {D}irac electron systems},\ }\href
  {https://doi.org/https://doi.org/10.1016/j.jpcs.2017.12.034} {\bibfield
  {journal} {\bibinfo  {journal} {J. Phys. Chem. Solids}\ }\textbf {\bibinfo
  {volume} {128}},\ \bibinfo {pages} {138} (\bibinfo {year}
  {2019})}\BibitemShut {NoStop}%
\bibitem [{\citenamefont {Chen}\ \emph {et~al.}(2017)\citenamefont {Chen},
  \citenamefont {Li}, \citenamefont {Zhu}, \citenamefont {Yang}, \citenamefont
  {Chen}, \citenamefont {Mao}, \citenamefont {Du}, \citenamefont {Wang},\ and\
  \citenamefont {Fang}}]{chen2017Bi}%
  \BibitemOpen
  \bibfield  {author} {\bibinfo {author} {\bibfnamefont {H.}~\bibnamefont
  {Chen}}, \bibinfo {author} {\bibfnamefont {L.}~\bibnamefont {Li}}, \bibinfo
  {author} {\bibfnamefont {Q.}~\bibnamefont {Zhu}}, \bibinfo {author}
  {\bibfnamefont {J.}~\bibnamefont {Yang}}, \bibinfo {author} {\bibfnamefont
  {B.}~\bibnamefont {Chen}}, \bibinfo {author} {\bibfnamefont {Q.}~\bibnamefont
  {Mao}}, \bibinfo {author} {\bibfnamefont {J.}~\bibnamefont {Du}}, \bibinfo
  {author} {\bibfnamefont {H.}~\bibnamefont {Wang}},\ and\ \bibinfo {author}
  {\bibfnamefont {M.}~\bibnamefont {Fang}},\ }\bibfield  {title} {\bibinfo
  {title} {Pressure induced superconductivity in the antiferromagnetic {D}irac
  material {B}a{M}n{B}i$_2$},\ }\href@noop {} {\bibfield  {journal} {\bibinfo
  {journal} {Sci. Rep.}\ }\textbf {\bibinfo {volume} {7}},\ \bibinfo {pages}
  {1} (\bibinfo {year} {2017})}\BibitemShut {NoStop}%
\bibitem [{SM()}]{SM}%
  \BibitemOpen
  \href@noop {} {\bibinfo  {journal} {See Supplemental Material
  https://journals.aps.org/ for experimental details with Refs. [28,31,34], url
  = {https://journals.aps.org/}}\ }\BibitemShut {NoStop}%
\bibitem [{\citenamefont {Wada}\ \emph {et~al.}(1984)\citenamefont {Wada},
  \citenamefont {Aoki},\ and\ \citenamefont {Fujita}}]{Wada}%
  \BibitemOpen
\bibfield  {journal} {  }\bibfield  {author} {\bibinfo {author} {\bibfnamefont
  {S.}~\bibnamefont {Wada}}, \bibinfo {author} {\bibfnamefont {R.}~\bibnamefont
  {Aoki}},\ and\ \bibinfo {author} {\bibfnamefont {O.}~\bibnamefont {Fujita}},\
  }\bibfield  {title} {\bibinfo {title} {{NMR} study of the electronic state
  and {P}eierls transitions in {N}b{S}e$_3$},\ }\href
  {https://doi.org/10.1088/0305-4608/14/6/019} {\bibfield  {journal} {\bibinfo
  {journal} {J. Phys. F: Met. Phys.}\ }\textbf {\bibinfo {volume} {14}},\
  \bibinfo {pages} {1515} (\bibinfo {year} {1984})}\BibitemShut {NoStop}%
\bibitem [{\citenamefont {Rahn}\ \emph {et~al.}(2017)\citenamefont {Rahn},
  \citenamefont {Princep}, \citenamefont {Piovano}, \citenamefont {Kulda},
  \citenamefont {Guo}, \citenamefont {Shi},\ and\ \citenamefont
  {Boothroyd}}]{Rahn2017}%
  \BibitemOpen
  \bibfield  {author} {\bibinfo {author} {\bibfnamefont {M.~C.}\ \bibnamefont
  {Rahn}}, \bibinfo {author} {\bibfnamefont {A.~J.}\ \bibnamefont {Princep}},
  \bibinfo {author} {\bibfnamefont {A.}~\bibnamefont {Piovano}}, \bibinfo
  {author} {\bibfnamefont {J.}~\bibnamefont {Kulda}}, \bibinfo {author}
  {\bibfnamefont {Y.~F.}\ \bibnamefont {Guo}}, \bibinfo {author} {\bibfnamefont
  {Y.~G.}\ \bibnamefont {Shi}},\ and\ \bibinfo {author} {\bibfnamefont {A.~T.}\
  \bibnamefont {Boothroyd}},\ }\bibfield  {title} {\bibinfo {title} {Spin
  dynamics in the antiferromagnetic phases of the dirac metals
  ${A}${M}n{B}i$_{2}$ (${A}=${S}r, {C}a)},\ }\href
  {https://doi.org/10.1103/PhysRevB.95.134405} {\bibfield  {journal} {\bibinfo
  {journal} {Phys. Rev. B}\ }\textbf {\bibinfo {volume} {95}},\ \bibinfo
  {pages} {134405} (\bibinfo {year} {2017})}\BibitemShut {NoStop}%
\bibitem [{\citenamefont {Liu}\ \emph {et~al.}(2017)\citenamefont {Liu},
  \citenamefont {Hu}, \citenamefont {Zhang}, \citenamefont {Graf},
  \citenamefont {Cao}, \citenamefont {Radmanesh}, \citenamefont {Adams},
  \citenamefont {Zhu}, \citenamefont {Cheng}, \citenamefont {Liu},
  \citenamefont {Phelan}, \citenamefont {Wei}, \citenamefont {Jaime},
  \citenamefont {Balakirev}, \citenamefont {Tennant}, \citenamefont {DiTusa},
  \citenamefont {Chiorescu}, \citenamefont {Spinu},\ and\ \citenamefont
  {Mao}}]{Liu2017}%
  \BibitemOpen
  \bibfield  {author} {\bibinfo {author} {\bibfnamefont {J.~Y.}\ \bibnamefont
  {Liu}}, \bibinfo {author} {\bibfnamefont {J.}~\bibnamefont {Hu}}, \bibinfo
  {author} {\bibfnamefont {Q.}~\bibnamefont {Zhang}}, \bibinfo {author}
  {\bibfnamefont {D.}~\bibnamefont {Graf}}, \bibinfo {author} {\bibfnamefont
  {H.~B.}\ \bibnamefont {Cao}}, \bibinfo {author} {\bibfnamefont {S.~M.~A.}\
  \bibnamefont {Radmanesh}}, \bibinfo {author} {\bibfnamefont {D.~J.}\
  \bibnamefont {Adams}}, \bibinfo {author} {\bibfnamefont {Y.~L.}\ \bibnamefont
  {Zhu}}, \bibinfo {author} {\bibfnamefont {G.~F.}\ \bibnamefont {Cheng}},
  \bibinfo {author} {\bibfnamefont {X.}~\bibnamefont {Liu}}, \bibinfo {author}
  {\bibfnamefont {W.~A.}\ \bibnamefont {Phelan}}, \bibinfo {author}
  {\bibfnamefont {J.}~\bibnamefont {Wei}}, \bibinfo {author} {\bibfnamefont
  {M.}~\bibnamefont {Jaime}}, \bibinfo {author} {\bibfnamefont
  {F.}~\bibnamefont {Balakirev}}, \bibinfo {author} {\bibfnamefont {D.~A.}\
  \bibnamefont {Tennant}}, \bibinfo {author} {\bibfnamefont {J.~F.}\
  \bibnamefont {DiTusa}}, \bibinfo {author} {\bibfnamefont {I.}~\bibnamefont
  {Chiorescu}}, \bibinfo {author} {\bibfnamefont {L.}~\bibnamefont {Spinu}},\
  and\ \bibinfo {author} {\bibfnamefont {Z.~Q.}\ \bibnamefont {Mao}},\
  }\bibfield  {title} {\bibinfo {title} {A magnetic topological semimetal
  {S}r$_{1-y}${M}n$_{1-z}${S}b$_2$ ($y$, $z <$ 0.1)},\ }\href
  {https://doi.org/10.1038/nmat4953} {\bibfield  {journal} {\bibinfo  {journal}
  {Nat. Mater.}\ }\textbf {\bibinfo {volume} {16}},\ \bibinfo {pages} {905}
  (\bibinfo {year} {2017})}\BibitemShut {NoStop}%
\bibitem [{\citenamefont {Beeman}\ and\ \citenamefont
  {Pincus}(1968)}]{Beeman1968}%
  \BibitemOpen
  \bibfield  {author} {\bibinfo {author} {\bibfnamefont {D.}~\bibnamefont
  {Beeman}}\ and\ \bibinfo {author} {\bibfnamefont {P.}~\bibnamefont
  {Pincus}},\ }\bibfield  {title} {\bibinfo {title} {Nuclear spin-lattice
  relaxation in magnetic insulators},\ }\href
  {https://doi.org/10.1103/PhysRev.166.359} {\bibfield  {journal} {\bibinfo
  {journal} {Phys. Rev.}\ }\textbf {\bibinfo {volume} {166}},\ \bibinfo {pages}
  {359} (\bibinfo {year} {1968})}\BibitemShut {NoStop}%
\bibitem [{\citenamefont {Sharapov}\ \emph {et~al.}(2004)\citenamefont
  {Sharapov}, \citenamefont {Gusynin},\ and\ \citenamefont
  {Beck}}]{sharapov2004Dirac}%
  \BibitemOpen
  \bibfield  {author} {\bibinfo {author} {\bibfnamefont {S.}~\bibnamefont
  {Sharapov}}, \bibinfo {author} {\bibfnamefont {V.}~\bibnamefont {Gusynin}},\
  and\ \bibinfo {author} {\bibfnamefont {H.}~\bibnamefont {Beck}},\ }\bibfield
  {title} {\bibinfo {title} {Magnetic oscillations in planar systems with the
  {D}irac-like spectrum of quasiparticle excitations},\ }\href@noop {}
  {\bibfield  {journal} {\bibinfo  {journal} {Phys. Rev. B}\ }\textbf {\bibinfo
  {volume} {69}},\ \bibinfo {pages} {075104} (\bibinfo {year}
  {2004})}\BibitemShut {NoStop}%
\bibitem [{\citenamefont {Tsaran}\ and\ \citenamefont
  {Sharapov}(2014)}]{Tsaran2014}%
  \BibitemOpen
  \bibfield  {author} {\bibinfo {author} {\bibfnamefont {V.~Y.}\ \bibnamefont
  {Tsaran}}\ and\ \bibinfo {author} {\bibfnamefont {S.~G.}\ \bibnamefont
  {Sharapov}},\ }\bibfield  {title} {\bibinfo {title} {Landau levels and
  magnetic oscillations in gapped {D}irac materials with intrinsic {R}ashba
  interaction},\ }\href {https://doi.org/10.1103/PhysRevB.90.205417} {\bibfield
   {journal} {\bibinfo  {journal} {Phys. Rev. B}\ }\textbf {\bibinfo {volume}
  {90}},\ \bibinfo {pages} {205417} (\bibinfo {year} {2014})}\BibitemShut
  {NoStop}%
\bibitem [{\citenamefont {Bridges}\ and\ \citenamefont
  {Clark}(1969)}]{Bridges1969}%
  \BibitemOpen
  \bibfield  {author} {\bibinfo {author} {\bibfnamefont {F.}~\bibnamefont
  {Bridges}}\ and\ \bibinfo {author} {\bibfnamefont {W.~G.}\ \bibnamefont
  {Clark}},\ }\bibfield  {title} {\bibinfo {title} {Quantum and other
  oscillations of the nuclear spin-lattice relaxation rate in
  $n\ensuremath{-}\mathrm{I}\mathrm{n}\mathrm{S}\mathrm{b}$},\ }\href
  {https://doi.org/10.1103/PhysRev.182.463} {\bibfield  {journal} {\bibinfo
  {journal} {Phys. Rev.}\ }\textbf {\bibinfo {volume} {182}},\ \bibinfo {pages}
  {463} (\bibinfo {year} {1969})}\BibitemShut {NoStop}%
\bibitem [{\citenamefont {Berg}\ \emph {et~al.}(1990)\citenamefont {Berg},
  \citenamefont {Dobers}, \citenamefont {Gerhardts},\ and\ \citenamefont
  {Klitzing}}]{Berg1990}%
  \BibitemOpen
  \bibfield  {author} {\bibinfo {author} {\bibfnamefont {A.}~\bibnamefont
  {Berg}}, \bibinfo {author} {\bibfnamefont {M.}~\bibnamefont {Dobers}},
  \bibinfo {author} {\bibfnamefont {R.~R.}\ \bibnamefont {Gerhardts}},\ and\
  \bibinfo {author} {\bibfnamefont {K.~v.}\ \bibnamefont {Klitzing}},\
  }\bibfield  {title} {\bibinfo {title} {Magnetoquantum oscillations of the
  nuclear-spin-lattice relaxation near a two-dimensional electron gas},\ }\href
  {https://doi.org/10.1103/PhysRevLett.64.2563} {\bibfield  {journal} {\bibinfo
   {journal} {Phys. Rev. Lett.}\ }\textbf {\bibinfo {volume} {64}},\ \bibinfo
  {pages} {2563} (\bibinfo {year} {1990})}\BibitemShut {NoStop}%
\bibitem [{\citenamefont {Fujii}\ \emph {et~al.}(2023)\citenamefont {Fujii},
  \citenamefont {Nakai}, \citenamefont {Hirata}, \citenamefont {Hasegawa},
  \citenamefont {Akahama}, \citenamefont {Ueda},\ and\ \citenamefont
  {Mito}}]{Fujii2023}%
  \BibitemOpen
  \bibfield  {author} {\bibinfo {author} {\bibfnamefont {T.}~\bibnamefont
  {Fujii}}, \bibinfo {author} {\bibfnamefont {Y.}~\bibnamefont {Nakai}},
  \bibinfo {author} {\bibfnamefont {M.}~\bibnamefont {Hirata}}, \bibinfo
  {author} {\bibfnamefont {Y.}~\bibnamefont {Hasegawa}}, \bibinfo {author}
  {\bibfnamefont {Y.}~\bibnamefont {Akahama}}, \bibinfo {author} {\bibfnamefont
  {K.}~\bibnamefont {Ueda}},\ and\ \bibinfo {author} {\bibfnamefont
  {T.}~\bibnamefont {Mito}},\ }\bibfield  {title} {\bibinfo {title} {Giant
  density of states enhancement driven by a zero-mode {L}andau level in
  semimetallic black phosphorus under pressure},\ }\href
  {https://doi.org/10.1103/PhysRevLett.130.076401} {\bibfield  {journal}
  {\bibinfo  {journal} {Phys. Rev. Lett.}\ }\textbf {\bibinfo {volume} {130}},\
  \bibinfo {pages} {076401} (\bibinfo {year} {2023})}\BibitemShut {NoStop}%
\end{thebibliography}%

\end{document}